\documentclass{aa}  

\usepackage{graphicx}
\usepackage{lscape}
\usepackage{natbib} 
\usepackage{siunitx}
\usepackage{url}
\usepackage{xcolor}

\usepackage[varg]{txfonts}

\bibpunct{(}{)}{;}{a}{}{,} 

\begin{document}

\title{The stellar and wind parameters of six prototypical HMXBs and their evolutionary status\thanks{Based on observations made with the NASA/ESA Hubble Space Telescope (HST), obtained from the data archive at the Space Telescope Science Institute. STScI is operated by the Association of Universities for Research in Astronomy, Inc. under NASA contract NAS 5-26555.}
}

   \author{R.~Hainich\inst{1}
          \and L.\,M.~Oskinova\inst{1}
          \and J.\,M.~Torrejón\inst{2}
          \and F.~Fuerst\inst{3}
          \and A.~Bodaghee\inst{4}
          \and T.~Shenar\inst{1,5}
          \and A.\,A.\,C. Sander\inst{1,6}
          \and H.~Todt\inst{1}
          \and K.~Spetzer\inst{4}
          \and W.-R.~Hamann\inst{1}
          }

   \institute{Institut f\"ur Physik und Astronomie,
              Universit\"at Potsdam,
              Karl-Liebknecht-Str. 24/25, D-14476 Potsdam, Germany \\
              \email{rhainich@astro.physik.uni-potsdam.de}
              \and 
              Instituto Universitario de F\`{\i}sica Aplicada a las Ciencias y las Tecnolog\`{\i}as, Universidad de Alicante, 03690 Alicante, Spain
              \and
              European Space Astronomy Centre (ESA/ESAC), 
              Science Operations Department, 
              Villanueva de la Cañada (Madrid), Spain
              \and
              Georgia College and State University, Dept. of Chemistry, Physics and Astronomy, 221 N. Wilkinson St., Milledgeville, GA 31061, USA
              \and
              Institute of astrophysics, KU Leuven, 
              Celestijnlaan 200D, 3001 Leuven, Belgium
              \and
              Armagh Observatory and Planetarium, 
              College Hill, Armagh, BT61 9DG, Northern Ireland
              }
   \date{Received <date> / Accepted <date>}


\abstract
%
%
{
High-mass X-ray binaries (HMXBs) are exceptional astrophysical laboratories that offer a rare glimpse into the physical processes that govern accretion on compact objects, massive-star winds, and stellar evolution. 
In a subset of the HMXBs, the compact objects accrete matter solely from winds of massive donor stars. 
These so-called wind-fed HMXBs are divided in persistent (classical) HMXBs and supergiant fast X-ray transients (SFXTs) according to their X-ray properties. While it has been suggested that this dichotomy depends on the characteristics of stellar winds, they have been poorly studied.
} 
%
%
{
With this investigation, we aim to remedy this situation by systematically analyzing donor stars of wind-fed HMXBs that are observable in the UV, concentrating on those with neutron star (NS) companions.
}
%
%
{
We obtained \textsl{Swift} X-ray data, HST UV spectra, and additional optical spectra for all our targets.
The spectral analysis of our program stars was carried out with the Potsdam Wolf-Rayet (PoWR) model atmosphere code. 
}
%
%
{
Our multi-wavelength approach allows us to provide stellar and wind parameters for six donor stars (four wind-fed systems and two OBe X-ray binaries). 
The wind properties are in line with the predictions of the line-driven wind theory. Based on the abundances, three of the donor stars are in an advanced evolutionary stage, while for some of the stars, the abundance pattern indicates that processed material might have been accreted. 
When passing by the NS in its tight orbit, the donor star wind has not yet reached its terminal velocity but it is still significantly slower; its speed is comparable with the orbital velocity of the NS companion.
There are no systematic differences between the two types of wind-fed HMXBs (persistent versus transients) with respect to the donor stars. 
For the SFXTs in our sample, the orbital eccentricity is decisive for their transient X-ray nature.
The dichotomy of wind-fed HMXBs studied in this work is primarily a result of the orbital configuration, while in general it is likely that it reflects a complex interplay between the donor-star parameters, the orbital configuration, and the NS properties. Based on the orbital parameters and the further evolution of the donor stars, the investigated HMXBs will presumably form Thorne–Żytkow objects in the future.
}
%
%
{}

\keywords{X-ray: binaries -- Binaries: close -- Stars: early type --
  Stars: atmospheres -- Stars: winds, outflows -- Stars: mass-loss}

\maketitle

\section{Introduction}
\label{sect:intro}

High-mass X-ray binaries (HMXBs) are binary systems consisting of a massive star, also denoted as the donor star, and a compact object, which is either a neutron star (NS) or a black hole (BH).  
These systems are characterized by high X-ray luminosities ($L_\mathrm{X} \approx 10^{36}\,\mathrm{erg/s}$) emitted by stellar material accreted onto the compact object.
Multi-wavelength studies of HMXBs offer the opportunity to contribute to a variety of physical and astrophysical research areas, including but not limited to accretion physics, stellar evolution, and the precursors of gravitational wave events. 

The HMXB population encompasses different types of binary systems. Depending on the orbital configuration, the evolution state of the donor star, and how the matter is channeled to the compact object, one can distinguish between different types: in Roche-lobe overflow (RLOF) systems, the compact object directly accretes matter via the inner Lagrangian point (L1). In wind-fed HMXBs, the compact object accretes from the wind of the donor star. In OBe X-ray binary systems, the donor stars are usually OB-type dwarfs with a decretion disk. The compact object in these kind of systems either accretes matter from these disks or from the donor-star winds. 

The wind-fed HMXBs are of particular interest (for recent reviews see \citealt{Martinez-Nunez2017} and \citealt{Sander2018b}). Here the compact object is situated in and accretes solely from the wind of the massive star, usually a supergiant. Therefore, these objects are also denoted as SgXBs. 
\citet{Ostriker1973} realized that accretion from a stellar wind onto a compact object is sufficient to power the high X-ray luminosities observed for these objects.
Depending on their X-ray properties, wind-fed HMXBs are distinguished into classical (or persistent) HMXBs and the so-called Supergiant Fast X-ray Transients (SFXTs). While the former always exhibit an X-ray luminosity on the order of $L_\mathrm{X} \approx 10^{36}\,\mathrm{erg/s}$, the latter are characterized by quiescent X-ray phases with $L_\mathrm{X} \approx 10^{32}-10^{34}\,\mathrm{erg/s}$, which are interrupted by sporadic X-ray flares ($L_\mathrm{X} \ge 10^{36}\,\mathrm{erg/s}$). 
Although the origin of this dichotomy is hitherto not understood, it is assumed that the donor stars play an important role in this picture \citep[e.g.,][]{intZand2007,Oskinova2012,Krticka2015,Gimenez-Garcia2016,Sidoli2018}. 

The Bondi-Hoyle-Lyttleton accretion mechanism \citep{Hoyle1939,Bondi1944} predicts that the X-ray luminosity ($L_\mathrm{X}$) of a wind-fed HMXB is very sensitive to the mass-loss rate ($\dot{M}$) and the wind velocity ($\vec{v}_\mathrm{wind}$) of the donor star,
\begin{equation}
\label{eq:lx}
L_\mathrm{X} \propto \dot{M}/v_\mathrm{rel}^4~,
\end{equation} 
where 
$v_\mathrm{rel} = |\,\vec{v}_\mathrm{wind} + \vec{v}_\mathrm{orb}\,|$ 
is the relative velocity of the wind matter captured by the compact object. 
The orbital velocity ($\vec{v}_\mathrm{orb}$) is often neglected while evaluating Eq.\,(\ref{eq:lx}), since stellar winds of OB-type stars have high terminal velocities, sometimes in excess of $2000\,\mathrm{km}\,\mathrm{s}^{-1}$. 
However, most HMXBs are compact systems with orbital periods of a few days \citep{Walter2015}. This implies that the distance between the compact objects and the donor stars are relatively small, which means that donor star winds will not have reached their terminal velocities at the position of the compact objects. On the other hand, the $v_\mathrm{orb}$ can be quite high, especially during periastron in eccentric systems. 
Therefore, it is of high importance to reliably quantify the role of $v_\mathrm{wind}$ and $v_\mathrm{orb}$ in these kind of systems, especially because of the strong dependence of $L_\mathrm{X}$ on $v_\mathrm{rel}$. 

The X-rays emitted by the compact object can, in turn, have a significant impact on the donor star's atmosphere and wind. These X-rays
strongly ionize a certain part of the donor star wind,
which can lead to significant changes in the observed spectra of these sources.
This is demonstrated by \citet{vanLoon2001} for important UV wind-lines using phase resolved spectroscopy of several donor stars.
The underlying mechanism is first discussed by \citet{Hatchett1977}, which is, therefore, also denoted as the Hatchett-McCray effect.
Depending on the wind density, the orbital configuration, and the amount of X-rays emitted by the compact object, its influence on the donor wind can be quite diverse
\citep[e.g.,][]{Blondin1990,Blondin1994}.

For high X-ray luminosities, \citet{Krticka2015} and \citet{Sander2018a} show that the donor wind velocity field in the direction of the compact object can be significantly altered. This is because the radiation of the compact object changes the ionization balance in the donor star wind, leading to a modification of the radiative acceleration of the wind matter. In extreme cases the donor star winds can be virtually stopped or even disrupted. 

In this work, we concentrate on wind-fed HMXBs with NS companions and moderate X-ray luminosities, where the Hatchett-McCray effect is of modest importance and the winds are not significantly disturbed. However, even for those systems, the X-rays need to be accounted for during the spectral analysis, since they might have a noticeable effect on the ionization balance in the donor star wind and consequently on the spectra.

Despite the strong connection between the X-ray properties of wind-fed HMXBs and the properties of the donor stars, only a few of these stars have been studied so far. 
One reason for this deficiency is that most of the wind-fed HMXBs are highly obscured. Therefore, the most important wavelength range for the analysis of OB-star winds, the UV that provides essential wind diagnostics, is often not accessible. 
In this work, we analyze four wind-fed HMXBs and two OBe X-ray binaries that are observable in the UV.

The paper is organized as follows: in Sect.\,\ref{sect:sample}, we introduce our sample, while the data used in this work are described in Sect.\,\ref{sect:data}. The atmosphere models and the fitting process are outlined in Sect.\,\ref{sect:modelling}. Our results are presented in Sect.\,\ref{sect:parameters}, and discussed in Sects.\,\ref{sect:evolution} and \ref{sect:x-rays}. The summary and conclusions can be found in Sect.\,\ref{sect:conclusions}. 
Additional tables, comments on the individual objects, and the complete spectral fits are presented in Appendices\,\ref{sec:addtables}, \ref{sec:comments}, and \ref{sect:spectra}, respectively.

\section{The sample}
\label{sect:sample}

\begin{table*}[htbp]
    \caption{Spectral classifications, distances, and common aliases} 
    \label{table:sample}
    \centering  
    \begin{tabular}{lllccl}
        \hline\hline 
        Name &
        HMXB type &
        Spectral type &
        Reference &
        Distance\tablefootmark{a} &
        Alias names  \rule[0mm]{0mm}{3.5mm} \\
        & & of donor & & (kpc) &  \\ 
        \hline  
        \object{HD\,153919}  &  persistent &  O6\,Iafpe           &  1 & $1.7^{+0.3}_{-0.2}$ & \object{4U\,1700-37}, \object{V*\,V884\,Sco} \rule[0mm]{0mm}{4.0mm}\\
        \object{BD+60\,73}   &  intermediate& BN0.7\,Ib            & 2 & $3.4^{+0.3}_{-0.2}$ & \object{IGR\,J00370+6122}   \rule[0mm]{0mm}{3.5mm}\\
        \object{LM\,Vel} &  SFXT       &  O8.5\,Ib-II(f)p      & 3 & $2.2^{+0.2}_{-0.1}$ & \object{HD\,74194}, 	\object{IGR\,J08408-4503}      \rule[0mm]{0mm}{3.5mm}\\
        \object{HD\,306414}  &  SFXT       &  B0.5\,Ia             & 4 & $6.5^{+1.4}_{-1.1}$ & \object{IGR\,J11215-5952}   \rule[0mm]{0mm}{3.5mm}\\
        \object{BD+53\,2790} &  persistent\,/\,Oe X-ray&  O9.5\,Vep      & 5 & $3.3^{+0.4}_{-0.3}$ & \object{4U\,2206+54}   \rule[0mm]{0mm}{3.5mm}\\
        \object{HD\,100199}  &  Be X-ray   &  B0\,IIIne            & 6 & $1.3^{+0.1}_{-0.1}$ & \object{IGR\,J11305-6256}   \rule[0mm]{0mm}{3.5mm}\\
        \hline 
    \end{tabular}
    \tablebib{
        (1) \citet{Sota2014};
        (2) \citet{Gonzalez-Galan2014};
        (3) \citet{Sota2014};
        (4) \citet{Lorenzo2014};
        (5) \citet{Blay2006};
        (6) \citet{Garrison1977};
    }
    \tablefoot{
           \tablefoottext{a}{The distances are adopted from \citet{Bailer-Jones2018}. These distances are based on the Gaia DR2 measurements \citep{Gaia2018} and were calculated by means of a Bayesian approach assuming an exponentially decreasing space density with the distance.}
    }
\end{table*}

While about 30 wind-fed HMXBs are known in our Galaxy (see \citealt{Martinez-Nunez2017} for a recent compilation), most of these objects are located in the Galactic plane \citep{Chaty2008}. Therefore, they are often highly obscured and are not observable in the UV. However, ultraviolet resonance lines allow to characterize even the relatively weak winds of B-type stars \citep[e.g.,][]{Prinja1989,Oskinova2011}. Since the determination of wind parameters is the main objective of this study, we restrict our sample to those wind-fed HMXBs that are observable in the UV.

In addition to Vela X-1, which has been analyzed previously by \citet{Gimenez-Garcia2016}, only four more wind-fed HMXBs meet the above condition, namely HD\,153919 (4U\,1700-37), BD+60\,73 (IGR\,J00370+6122), LM\,Vel (IGR\,J08408-4503), and HD\,306414 (IGR\,J11215-5952).
The latter two systems are SFXTs, while the first one is a persistent HMXB, and BD+60\,73 (IGR\,J00370+6122) exhibits properties of both types. 
Our sample also includes the Be X-ray binary HD\,100199 (IGR\,J11305-6256) and BD+53\,2790 (4U\,2206+54), which is classified as an Oe X-ray binary or as a persistent wind-fed binary with a non evolved donor. The latter classification is based on its X-ray properties, while the former is a result of the prominent hydrogen emission lines that are visible in optical spectra of this object. These lines are most likely formed in a decretion disk of the donor star. Thus, this systems might actually be intermediate between the classical wind-fed HMXB and the OBe X-ray binaries.
The HMXB type, the spectral classification of the donor, and common alias names of the investigated systems are given in Table\,\ref{table:sample}. 

The orbital parameters of the investigated systems and the spin period of the neutron stars are compiled from the literature and listed in Table\,\ref{table:orbital}. The only exception is HD\,100199 because neither the orbit nor the properties of its NS are known. 

\section{The data}
\label{sect:data}

\begin{table*}[htbp]
    \caption{Orbital parameters and the spin period of the neutron star} 
    \label{table:orbital}
    \centering  
    \begin{tabular}{llclcllSc}
        \hline\hline 
        Identifier &
        \multicolumn{1}{c}{Orbital period} &
        Ref. &
        \multicolumn{1}{l}{Eccentricity} &
        Ref. &
        \multicolumn{1}{c}{$T_0$}
        & Ref. &
        \multicolumn{1}{c}{Spin period} &
        Ref. \rule[0mm]{0mm}{3.5mm} \\
        & \multicolumn{1}{c}{(d)} & & & & \multicolumn{1}{c}{(MJD)} &  \multicolumn{1}{c}{(s)} & &\\
        \hline  
        HD\,153919  & $3.411660 \pm 0.000004$ & 1 & $0.008-0.22$ & 1,2 & $49149.412 \pm 0.006$ & 1 & \multicolumn{1}{c}{-} &  \rule[0mm]{0mm}{4.0mm}\\
        BD+60\,73   & $15.661 \pm 0.0017$ & 3 & $0.56 \pm 0.07$ & 3 & $55084.0  \quad\pm 0.4$ & 3 & 346.0 & 4 \\
        LM\,Vel & $9.5436 \pm 0.0002$  & 5 & $0.63 \pm 0.03$ & 5 & $54634.45 \,\,\,\pm 0.04$ & 5 & \multicolumn{1}{c}{-} &  \\
        HD\,306414  & $\sim164.6$ & 6 & $\sim0.8$\tablefootmark{a} & 7 & \multicolumn{1}{c}{-} & & 186.78 & 6,8\\
        BD+53\,2790 & $\sim 9.568$   & 9,10,11,12 & $0.30 \pm 0.02$ & 12 & \multicolumn{1}{c}{-} & & 5750. & 13 \\
        \hline 
    \end{tabular}
    \tablebib{(1)~\citet{Islam2016};
        (2) \citet{Hammerschlag-Hensberge2003};
        (3) \citet{Gonzalez-Galan2014};
        (4) \citet{intZand2007};
        (5) \citet{Gamen2015};;
        (6) \citet{Romano2009};
        (7) \citet{Lorenzo2014};
        (8) \citet{Swank2007};
        (9) \citet{Corbet2001}
        (10) \citet{Ribo2006};
        (11) \citet{Reig2009};
        (12) \citet{Stoyanov2014};
        (13) \citet{Torrejon2018}
    }
    \tablefoot{
        \tablefoottext{a}{uncertain}
    }
\end{table*}

\subsection{Spectroscopy}

For our UV survey of wind-fed HMXBs, we made use of the \emph{Space Telescope Imaging Spectrograph} \citep[STIS,][]{Woodgate1998,Kimble1998} aboard the \textsl{HST}. These high resolution, high S/N spectra (Proposal ID: 13703, PI: L.\,M. Oskinova) cover important wind diagnostics in the range 1150-1700\,\AA. In this paper, we use the automatically reduced data that are provided by the \textsl{HST} archive. For three of our program stars, far UV data obtained with the \emph{Far Ultraviolet Spectroscopic Explorer} \citep[FUSE,][]{Moos2000} were retrieved from the MAST archive. 

These data are complemented by optical spectroscopy from different sources. For HD\,100199, HD\,306414, LM\,Vel, and HD\,153919, we use data taken with the \emph{Fiber-fed Extended Range Optical Spectrograph} \citep[FEROS,][]{Kaufer1999} mounted at the 2.2\,m telescope operated at the European Southern Observatory (ESO) in La Silla. These data sets were downloaded from the ESO archive. From the same repository, we also retrieved \emph{FOcal Reducer and low dispersion Spectrograph} \citep[FORS,][]{Appenzeller1998} spectra for HD\,153919. Optical spectra for BD+60\,73 were kindly provided by A.\,González-Galán. These spectra were taken with the \emph{high-resolution FIbre-fed Echelle Spectrograph} \citep[FIES,][]{Telting2014} mounted on the Nordic Optical Telescope (NOT) and published in \citet{Gonzalez-Galan2014}. For BD+53\,2790, we downloaded a low resolution spectrum from the VizieR archive that was taken by \citet{Munari2002} with a \emph{Boller \& Chivens Spectrograph} of the Asiago observatory. In addition, we obtained an optical spectrum of BD+53\,2790 with a \emph{DADOS spectrograph} in combination with two different SBIG cameras (SFT8300M \& ST-8XME) mounted to the Overwhelmingly Small Telescope (OST) of the student observatory at the University of Potsdam. Default data reduction steps were performed for this data set using calibration data (dome flats, dark frames, \element{Hg}\element{Ar}-lamp spectrum) taken immediately after the science exposures. 
Finally Near-IR spectroscopy was obtained during the night of 2014 September 1, using the \emph{Near Infrared Camera and Spectrograph} (NICS) mounted at the 3.5-m Telescopio Nazionale Galileo (TNG) telescope (La Palma island). Medium-resolution spectra (3.5 \AA pixel$^{-1}$) were taken with the $H$ and $K_{\rm b}$ grisms under good seeing conditions. Details on the reduction process can be seen in \citet{Rodes-Roca2018}.
The individual spectral exposures used in this work are listed in Table\,\ref{table:spectra}. 
In this Table, we also give the phase at which the observations were taken for those systems where ephemerides are available (see Table\,\ref{table:orbital} and references therein).

\subsection{Photometry}

We compiled $UBVRI$ photometry from various sources \citep{Zacharias2004b,DENIS2005,Mermilliod2006,Anderson2012,Reig2015} for all our program stars. G-band photometry was retrieved from the Gaia DR1 release \citep{Gaia2016}. 
Near-infrared photometry ($J, H, K_S$) was obtained from \citet{Cutri2003}, while WISE photometry is available from \citet{Cutri2012a} for all our targets. Moreover, we made use of MSX infrared photometry \citep{Egan2003} for HD\,153919. The complete list of photometric measurements used for the individual objects is compiled in Table\,\ref{table:photometry}.

\subsection{X-ray data}
\label{subsect:xray_data}

For all our \textsl{HST} observations, we obtained quasi-simultaneous X-ray data with the Neil Gehrels Swift Observatory \citep[\textsl{Swift},][]{Gehrels2004}. In addition, strictly simultaneous {\em Chandra} X-ray and \textsl{HST} UV observations were performed for  \object{HD\,153919} ({\em Chandra} ObsID.\,17630, exposure time 14.6\,ks). 

The data obtained with the X-ray telescope \citep[XRT,][]{Burrows2005} aboard \textsl{Swift} are reduced using the standard XRT pipeline as part of HEASOFT v6.23. To extract the source spectra from data gathered while the XRT was in the photon counting (PC) mode, we used a circular region centered at its J2000 coordinates with a 25\,\arcsec\ radius or 80\,\arcsec\ radius depending on the source characteristics. Background counts were extracted from an annulus encompassing the source extraction region. When XRT was in the window timing mode (WT), the source extraction region consisted of a square with a width of 40 pixels while background counts were extracted from a similar-sized region situated away from the source.

The observed spectra were fitted using a suit of various X-ray spectral fitting software packages. For all objects, the photoionization cross-sections from \citet{Verner1996} and abundances from \citet{Wilms2000} were employed. The goal of X-ray spectral fitting was to provide the parameters describing the X-ray radiation field in the format required for the stellar atmosphere modeling (see Sect.\,\ref{sect:models}). X-ray spectra of HMXBs are typically well represented by power law spectral models,  which are not yet implemented in our stellar atmosphere model. Therefore, we decided to fit the observed spectra using a fiducial black body spectral model. The fitting returns a ``temperature'' parameter $T_\mathrm{X}$, which is not-physical but is employed to describe the spectral hardness and X-ray photon flux.     

The \textsl{Swift} XRT observation of \object{HD\,153919} was taken in the WT mode. We extracted 43,640 net source counts during 5270\,s of exposure time. After rebinning the spectral data to contain a minimum of 20 net counts per bin, we fit an absorbed ($N_{\mathrm{H}} = 15_{-2}^{+4} \times 10^{22}$\,cm$^{-2}$) blackbody ($k_{\mathrm{B}}T = 2.1\pm0.1$\,keV) plus a power-law component ($\Gamma = 4 \pm 1$). The observed X-ray
flux is $1.2_{-0.1}^{+0.2}\times 10^{-9}$\,erg\,cm$^{-2}$\,s$^{-1}$.  {\em Chandra} observations of \object{HD\,153919} are presented by \citet{Martinez-Chicharro2018}. Towards the end of the observation which lasted about 4\,h, the source experienced a flare with X-ray flux increasing  by a factor of three. Our \textsl{HST} observations were partially obtained during the end of this flare.  

Nineteen source ($+$background) counts were gathered during the XRT observation of \object{BD+60\,73} taken on the same day as the HST observation (ObsID 00032620025). Without rebinning the data, and assuming C-statistics \citep{Cash1979}, we fit an absorbed blackbody model and obtained spectral parameters that were poorly constrained ($N_{\mathrm{H}} = 4_{-3}^{+28}\times 10^{22}$\,cm$^{-2}$ and $k_{\mathrm{B}}T \sim 1$\,keV) with an observed flux of $6.2\times 10^{-14}$\,erg\,cm$^{-2}$\,s$^{-1}$.

HD\,306414 was not detected in any of the contemporaneous \textsl{Swift} observations, and therefore we could not measure its X-ray flux. HD\,100199 was marginally detected with $12\pm4$ photons in an observation one day before the \textsl{HST} observation (ObsID 00035224007). We estimate a flux of $2.7^{+3.9}_{-1.8}\times10^{-13}\,\mathrm{erg}\,\mathrm{cm}^{-2}\,\mathrm{s}^{-1}$ from these data.

LM\,Vel was also very X-ray faint during the \textsl{HST} observation (ObsID 00037881107). We therefore use \textsl{Swift} data taken a few days earlier (ObsID 00037881103) to measure the spectral shape. We find that a thermal blackbody model describes the data well, and use this model to fit the simultaneous data. There we find a flux of $5.8\times10^{-13}\,\mathrm{erg}\,\mathrm{cm}^{-2}\,\mathrm{s}^{-1}$ between
3--10\,keV.

For \object{BD+53\,2790} (\object{4U\,2206+543}), we extracted 2511 net source counts during an 1106\,s XRT observation in WT mode. The spectral data were arranged in order to contain at least 20 counts per bin, and were then fit with an absorbed blackbody model ($N_{\mathrm{H}} \leq 8\times 10^{21}$\,cm$^{-2}$ and $k_{\mathrm{B}}T = 1.3 \pm 0.1 $\,keV). The model derived flux is $1.3\pm0.1 \times 10^{-11}$\,erg\,cm$^{-2}$\,s$^{-1}$.

A compilation of the X-ray data used in this work can be found in Table\,\ref{table:xray_data}, while the derived X-ray luminosities are listed in Table\,\ref{table:Lxray}.

\begin{table}[htbp]
    \caption{X-ray luminosities measured at times close to the \textsl{HST} observations (see text for details)} 
    \label{table:Lxray}
    \centering  
    \begin{tabular}{llc}
        \hline\hline 
        \multicolumn{2}{c}{Identifier} & 
        $\log L_\mathrm{X}$\,[erg/s] \rule[0mm]{0mm}{3.5mm} \\
        \hline  
        HD\,153919     & 4U\,1700-37      &  36.03  \rule[0mm]{0mm}{4.0mm}\\
        BD+60\,73      & IGR\,J00370+6122 &  31.90  \\
        LM\,Vel        & IGR\,J08408-4503 &  32.50  \\
        HD\,306414     & IGR\,J11215-5952 &  \multicolumn{1}{c}{--}  \\
        BD+53\,2790    & 4U\,2206+54      &  34.24  \\
        HD\,100199     & IGR\,J11305-6256 & \multicolumn{1}{c}{--}   \\
        \hline 
    \end{tabular}
    \tablefoot{
        {A -- indicates that the source was below the detection limit of \textsl{Swift} during the observations.}
    }
\end{table}

\section{Spectral modeling}
\label{sect:modelling}

\subsection{Stellar atmosphere models}
\label{sect:models}

The spectral analyses presented in this paper were carried out with the Potsdam Wolf-Rayet (PoWR) models. PoWR is a state-of-the-art code for expanding stellar atmospheres. The main assumption of this code is a spherically symmetric outflow. The code accounts for deviation from the local dynamical equilibrium (non-LTE), iron line blanketing, wind inhomogeneities, a consistent stratification in the quasi hydrostatic part, and optionally also for irradiation by X-rays. The rate equations for the statistical equilibrium are solved simultaneously with the radiative transfer in the comoving frame, while energy conservation is ensured. Details on the code can be found in \citet{Graefener2002}, \citet{Hamann2003}, \citet{Todt2015}, and \citet{Sander2015}.

The inner boundary of the models is set to a Rosseland continuum optical depth $\tau_\mathrm{ross}$ of 20, defining the stellar radius $R_*$. The stellar temperature $T_*$ is the effective temperature that corresponds to $R_*$ via the Stefan-Boltzmann law,
\begin{equation}
\label{eq:sblaw}
L = 4 \pi \sigma_\mathrm{SB} R_\ast^2 T_\ast^4~\,,
\end{equation} 
with $L$ being the luminosity. The outer boundary is set to $R_\mathrm{max} = 100\,R_*$, which proved to be sufficient for our program stars. 

In the subsonic part of the stellar atmosphere, the velocity field $v(r)$ is calculated consistently such that the quasi-hydrostatic density stratification is fulfilled. In the wind, corresponding to the supersonic part of the atmosphere, a $\beta$-law \citep{Castor1979,Pauldrach1986} is assumed.
A double-$\beta$ law \citep{Hillier1999,Graefener2005} in the form described by \citet{Todt2015} is used for those objects where $\beta$ values larger than unity are required to achieve detailed fits. For the first exponent we always assume 0.8, while the second exponent is adjusted during the spectral fitting procedure. 
The gradient of such a double-$\beta$ law is steeper at the bottom of the wind than for a single $\beta$-law with a large exponent. 

In the main iteration, line broadening due to natural broadening, thermal broadening, pressure broadening, neglected multiplet splitting, and turbulence is approximately accounted for by assuming Gaussian line profiles with a Doppler width of $30\,\mathrm{km}\,\mathrm{s}^{-1}$. The turbulent pressure is accounted for in the quasi hydrostatic equation (see \citealt{Sander2015} for details). In the formal integral, line broadening is treated in all detail. For the microturbulence we set $\xi = 10\,\mathrm{km}\,\mathrm{s}^{-1}$ in the photosphere, growing proportional with the wind velocity up to a value of $\xi(R_\mathrm{max}) = 0.1\,v_\infty$.
The only exceptions are the supergiants HD\,306414 and BD+60\,73 where higher  $\xi$ values are necessary to reproduce the observation (see Appendix\,\ref{sec:comments} for details).
The atmospheric structures (e.g., the density and the velocity stratification) of the final models for the donor stars are listed in Tables\,\ref{table:struct_hd153919} -- \ref{table:struct_hd100199}.

Wind inhomogeneities are accounted for in the ``microclumping'' approach that assumes optically thin clumps \citep{Hillier1991,Hamann1998}. The density contrast between the clumps of an inhomogeneous model and a homogeneous one (with the same mass-loss rate $\dot{M}$) is described by the clumping factor $D$. Since the interclump medium is assumed to be void, $D$ is the inverse of the clump's volume filling factor $f_\mathrm{V} = D^{-1}$. According to hydrodynamical simulations \citep[e.g.,][]{Runacres2002,Sundqvist2018}, a radial dependency is expected for the clumping factor. Here, we use the clumping prescription suggested by \citet{Martins2009}. The clumping onset (parameterized by $v_\mathrm{cl}$), where the clumping becomes significant, is set to 10\,$\mathrm{km}\,\mathrm{s}^{-1}$, since this results in the best fits for all objects where this property could be constrained. The clumping factor is adjusted for each individual object. 

The PoWR code accounts for ionization due to X-rays. The X-ray emission is modeled as described by \citet{Baum1992}, assuming that the only contribution to the X-ray flux is coming from free-free transitions. Since the current generation of PoWR models is limited to spherical symmetry, the X-rays are assumed to arise from an  optically-thin spherical shell around the star. 
The X-ray emission is specified by three free parameters, which are the fiducial temperature of the X-ray emitting plasma $T_\mathrm{X}$, the onset radius of the X-ray emission $R_0$ ($R_0 > R_\ast$), and a filling factor $X_\mathrm{fill}$, describing the ratio of shocked to non-shocked plasma. For our HMXBs, the onset radius is set to the orbital distance between the donor star and the NS companion. 
The temperature of the X-ray emitting plasma are obtained from fits of the observed X-ray spectra (see Sect.\,\ref{subsect:xray_data}). The X-ray filling factor is adjusted such that the wavelength integrated X-ray flux from the observations is reproduced by the model.

\begin{figure}[tbp]
    \centering
    \includegraphics[angle=90,width=\hsize]{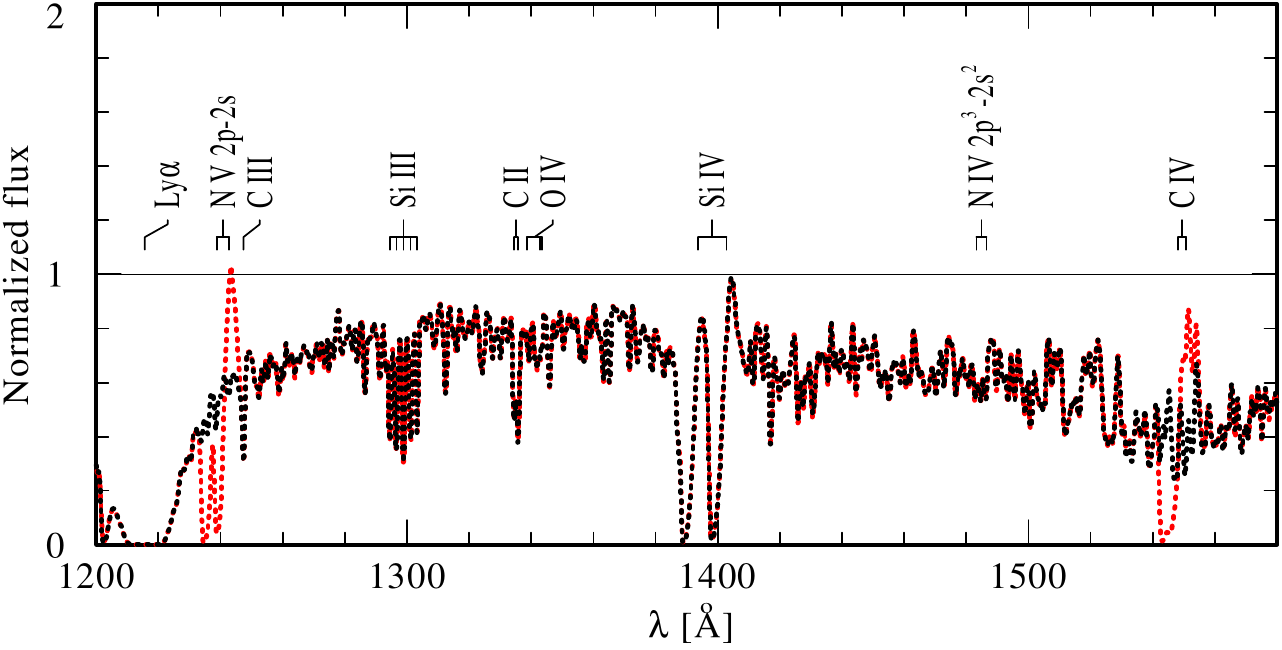}
    \caption{
        Comparison between two model spectra calculated for BD+60\,73 to illustrate the effect of the X-rays (red: with X-rays, black: without X-rays).
    }
    \label{fig:xray_comp}
\end{figure}

The effects of the X-ray field on the emergent spectra are illustrated in Fig.\,\ref{fig:xray_comp}. While the photospheric absorption lines are not affected at all, certain wind lines change significantly. Whether the lines become stronger or weaker depends on the individual combination of the wind density at the position of the NS, the ionization balance in the wind, and the hardness and intensity of the X-rays injected. 
There is some parameter degeneracy as, for some models, nearly identical line profiles are obtained when reducing $\dot{M}$ and instead increasing the X-ray filling factor. 
Fortunately, this ambiguity can be avoided in the analysis of most of our targets because  the X-ray field is constrained from observations (see Sect.\,\ref{subsect:xray_data}). 
The injected X-ray radiation is often needed to reproduce the wind lines in the UV and, hence, to measure the terminal wind velocity and mass-loss rate.

Complex model atoms of \element{H}, \element{He}, \element{C}, \element{N}, \element{O}, \element{Mg}, \element{Si}, \element{P}, and \element{S} (see Table\,\ref{table:model_atoms} for details) are considered in the non-LTE calculations. The multitude of levels and line transitions inherent to the iron group elements (\element{Sc}, \element{Ti}, \element{V}, \element{Cr}, \element{Mn}, \element{Fe}, \element{Co}, and \element{Ni}) are treated in a superlevel approach \citep[see\ ][]{Graefener2002}.

\subsection{Applicability of the models}
One of the main assumption of the PoWR models as well as all other stellar atmosphere codes, with the exception of the PHOENIX/3D code \citep{Hauschildt2014}, is spherical symmetry. 
In HMXBs, however, the spherical symmetry is broken by the presence of the compact object. On the other hand, the X-ray luminosities are often quite modest in HMXBs with NS companions. This is also the case for the systems studied in this work, as illustrated by the values given in Table\,\ref{table:Lxray}. For all but one sources, the X-ray luminosities are below $\log L_\mathrm{X} = 35$\,[erg/s]. 

\begin{figure*}[tbp]
    \centering
    \includegraphics[angle=90,width=\hsize]{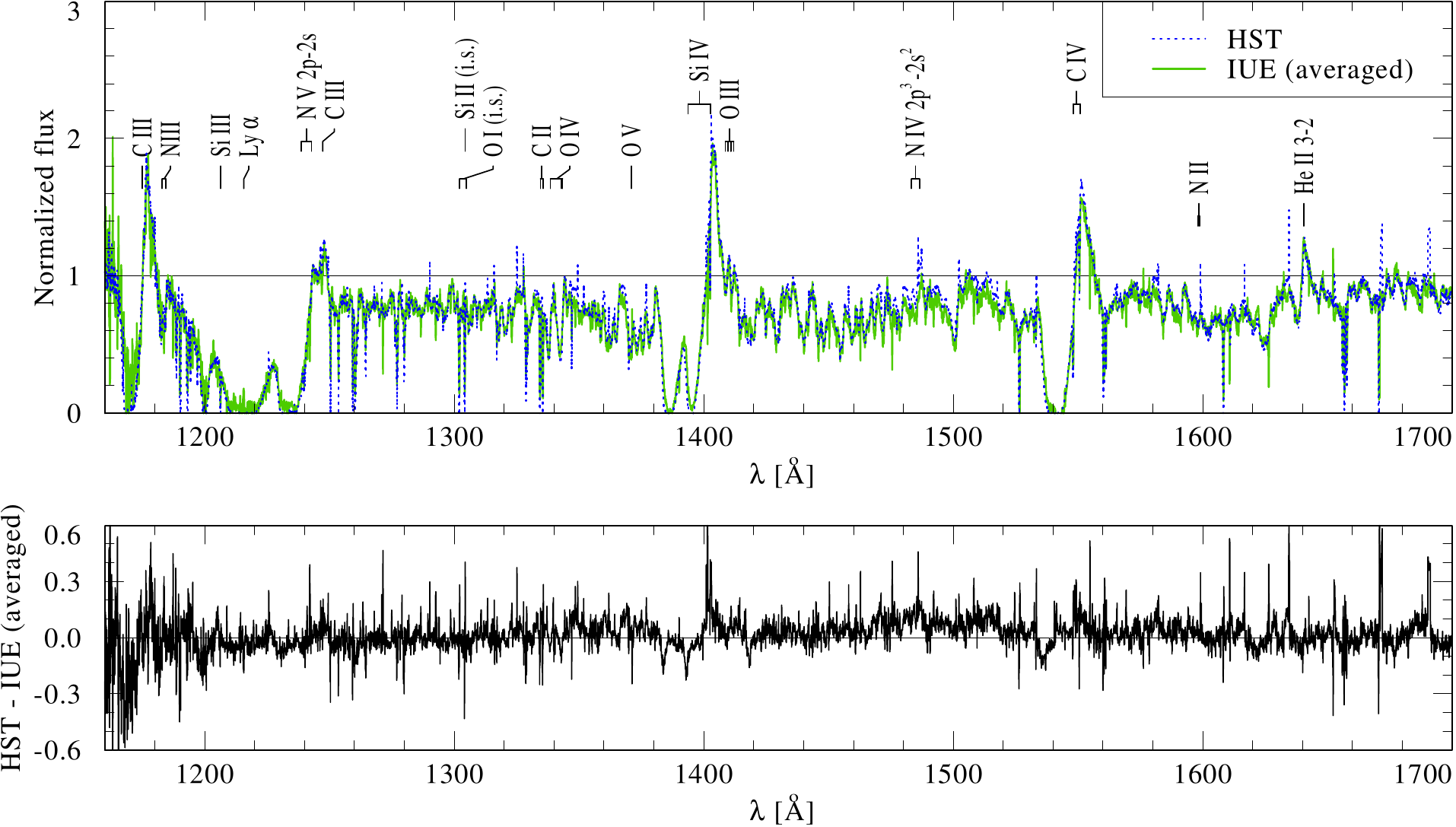}
    \caption{
        HD\,153919: comparison between our \textsl{HST} spectrum and an averaged \textsl{IUE} spectrum constructed from data taken between 1978 and 1989. Upper panel: \textsl{HST} spectrum (blue dotted line) and \textsl{IUE} spectrum (green continuous line); lower panel: difference between the \textsl{HST} spectrum and the averaged \textsl{IUE} spectrum
    }
    \label{fig:hd153919-iue}
\end{figure*}

For those X-ray luminosities, we expect that the disruptive effect of the X-rays emitted by the NS on the donor star wind is relatively limited. The only exception might be HD\,153919 (4U\,1700-37) that exhibited an X-ray luminosity of $\log L_\mathrm{X} = 36.03$\,[erg/s] during our \textsl{HST} observation. The \textsl{HST} spectrum of this source was taken during the end of an X-ray outburst described in 
\citet{Martinez-Chicharro2018}. However, only minor variations are present in the \textsl{HST} spectrum compared to earlier observations with the \textsl{IUE} satellite. 
This is illustrated in Fig.\,\ref{fig:hd153919-iue}, where we compare our \textsl{HST} spectrum with an averaged \textsl{IUE} spectrum constructed from observations in the high resolution mode that were taken between 1978 and 1989 with the large aperture. We used all available data sets with the exception of one exposure (Data ID: SWP36947) that exhibits a significantly lower flux compared to all other observations. 
The wind of HD\,153919 does not show any sign of inhibition, suggesting that the volume significantly affected by the X-ray emission of the NS is rather small. This is consistent with the findings by \citet{vanLoon2001}.

This gives us confidence that the winds of the donor stars in the studied systems are not disrupted by the X-ray emission of the NSs and that the applied models are valid within their limitations. However, for the individual objects, observational time series are necessary to confirm this.

\subsection{Spectral analysis}

\begin{figure*}[tbp]
    \centering
    \includegraphics[angle=90,width=\hsize]{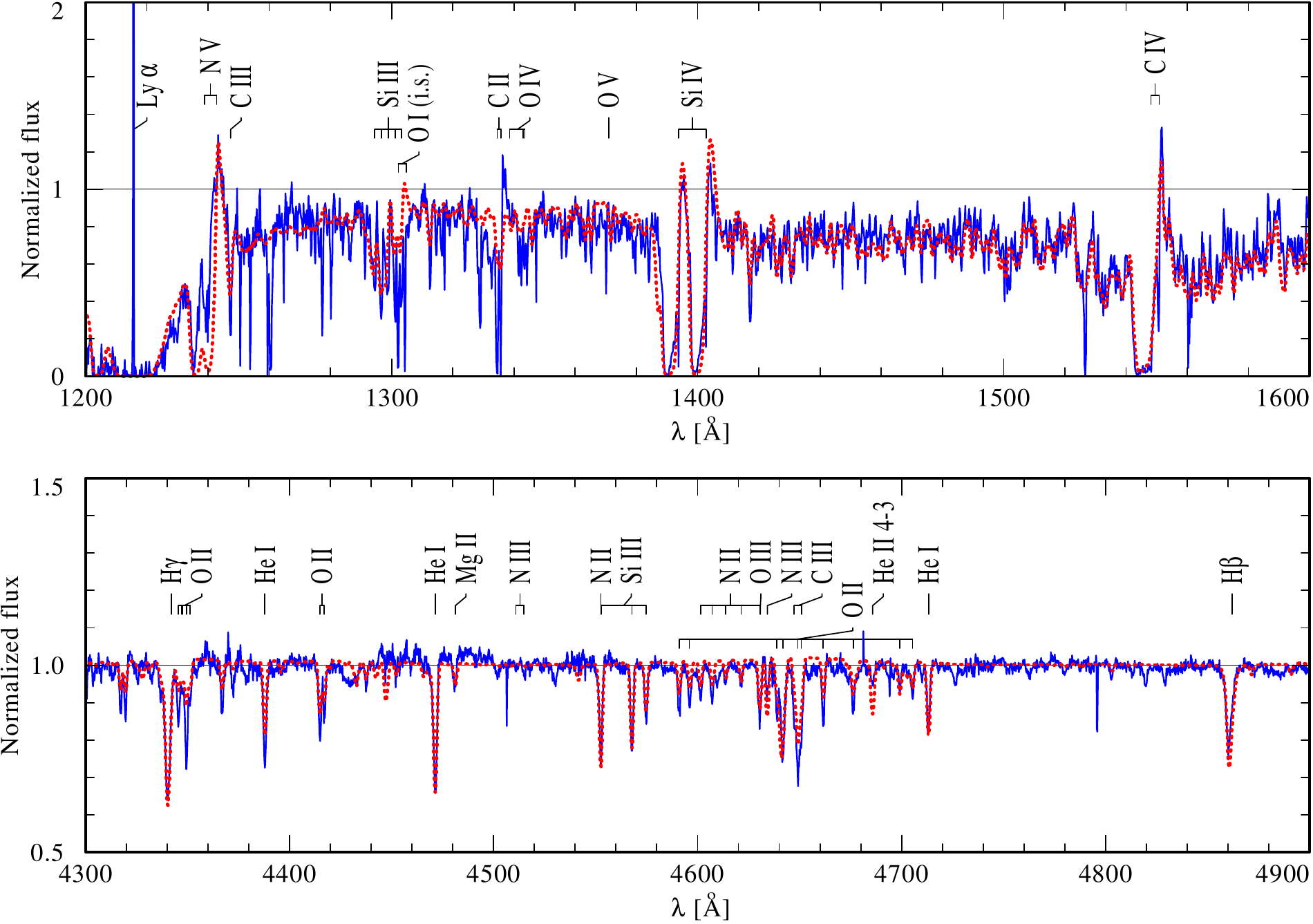}
    \caption{
        Normalized line spectrum of HD\,306414. The upper panel depicts a part of the UV spectrum, while the lower panel shows a section of the optical spectrum. The observation is shown in blue. The best fitting synthetic spectrum is overplotted by a dashed red line. 
    }
    \label{fig:spec}
\end{figure*}

An in-depth spectral analysis of a massive star with non-LTE model spectra is an iterative process. 
Our goal is to achieve a overall best model fit to the observed data, while weighting the diagnostics according to their sensibility to the stellar parameters as described below. 
Starting from an estimate of the stellar parameters based on the spectral type of the target, a first stellar atmosphere model is calculated and its emergent spectrum is compared to the observations. 
This and the subsequent comparisons are performed “by eye” without any automatic minimization procedures.
Based on the outcome of the initial comparison, the model parameters are adjusted, and a new atmosphere model is calculated. This procedure is repeated until satisfactory fits of the observations with the normalized line spectrum and the spectral energy distribution (SED) is achieved. As an example, the final fit of the normalized line spectrum of HD\,306414 is presented in Fig.\,\ref{fig:spec}.

For those objects in our sample with $T_\ast > 30\,\mathrm{kK}$, the stellar temperature is primarily derived from the equivalent-width ratio between \ion{He}{i} and \ion{He}{ii} lines, such as \ion{He}{i}\,$\lambda\lambda$\,4026, 4144, 4388, 4713, 4922, 5015, 6678 and \ion{He}{ii}\,$\lambda\lambda$\,4200, 4542, 5412, 6683. For stars with lower stellar temperatures, we additionally used the line ratios of \ion{Si}{iii} to \ion{Si}{iv} (\ion{Si}{iii}\,$\lambda\lambda$\,4553, 4568, 4575, 5740 and \ion{Si}{iv}\,$\lambda\lambda$\,4089, 4116) and \ion{N}{ii} to \ion{N}{iii} (\ion{N}{ii}\,$\lambda\lambda$\,4237, 4242, 5667, 5676, 5680, 5686, and \ion{N}{iii}\,$\lambda\lambda$\,4035, 4097).

The surface gravity $\log g_\mathrm{grav}$ is derived from the pressure broadened wings of the Balmer lines, focusing on the \element{H}$\gamma$ and \element{H}$\delta$ line, since \element{H}$\beta$ and \element{H}$\alpha$ are often affected by emission lines from the stellar wind.

\begin{figure*}[tbp]
    \centering
    \includegraphics[width=\hsize]{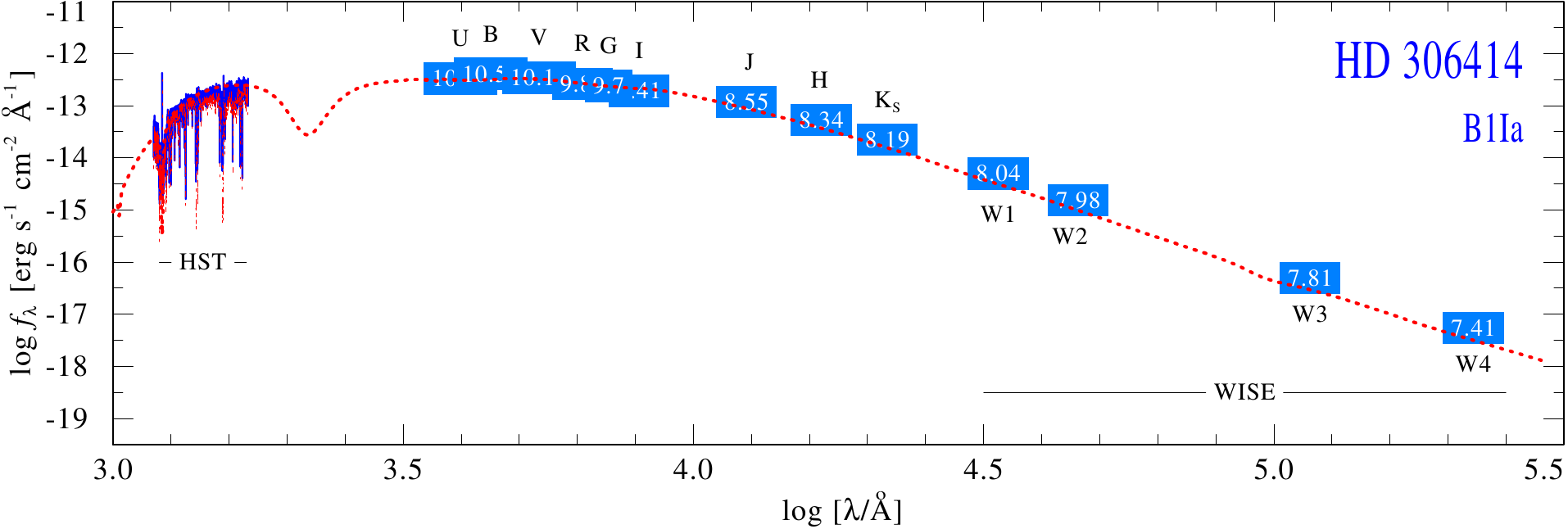}
    \caption{
        Spectral energy distribution (SED) of HD\,306414, composed of the flux calibrated \textsl{HST} spectrum (blue continuous line) and photometry (blue boxes, labeled with the corresponding magnitudes). The best fitting model SED is depicted by a dashed red line. The model flux is corrected for interstellar extinction and the geometric dilution according to the distance to HD\,306414 (6.5\,kpc). 
    }
    \label{fig:sed}
\end{figure*}

The luminosities of all our targets together with the color excess $E_{B-V}$ and the extinction-law parameter $R_V$ for the individual lines of sight are obtained from a fit of the corresponding model SED to photometry and flux calibrated spectra. For this purpose, different reddening laws are applied to the synthetic SEDs. The finally adopted reddening prescriptions are given in Table\,\ref{table:parameters}. Moreover, the model flux is scaled to the distance of the corresponding star, using the values compiled in Table\,\ref{table:sample}. For example, the SED fit of HD\,306414 is shown in Fig.\,\ref{fig:sed}. 

The projected rotational velocity and the microturbulence velocity in the photosphere are derived from the line profiles and the equivalent width of metal lines, such as \ion{Si}{iv}\,$\lambda\lambda$\,4089, 4116; \ion{Si}{iii}\,$\lambda\lambda$\,4553, 4568, 4575; \ion{Mg}{ii}\,4481\,$\lambda$; and \ion{C}{iv}\,$\lambda\lambda$\,5801, 5812. 
Macroturbulence is not considered in this approach, and thus the $v \sin i$ values reported in Table\,4 must be considered as upper limits. With the \texttt{iacob-broad} tool \citep{Simon-Diaz2014}, which separately determines a possible macroturbulent contribution to the line broadening, we obtained similar $v \sin i$ values within their error margins.
The terminal wind velocity and the radial dependence of the microturbulence velocity are simultaneously estimated from the extend and shape of the P Cygni absorption troughs of the UV resonance lines.
The $\beta$ parameter of the velocity law is adjusted such that the synthetic spectrum can reproduce the profiles of the UV resonance lines and the full-width at half-maximum (FWHM) of the H$\alpha$ emission. 
For the objects presented in this work, a double-$\beta$ law with a second $\beta$ exponent in the range of 1.2--3.0 result in slightly better spectral fits compared to the canonical $\beta$-law with $\beta = 0.8$ for O-type stars \citep{Kudritzki1989,Puls1996}.
Note that the mass-loss rate derived from a spectral fit also slightly depend on the used $\beta$ value.

The mass-loss rate and the clumping parameters are derived by fitting the wind lines in the UV and the optical. The main diagnostics for determining the mass-loss rates are the UV resonance lines exhibiting P Cygni line profiles, namely,  \ion{C}{iv}\,$\lambda\lambda$\,1548, 1551 and \ion{Si}{iv}\,$\lambda\lambda$\,1394, 1403. The clumping factor and the onset of the clumping are adjusted such that a consistent fit of unsaturated UV lines and H$\alpha$ could be achieved, utilizing the different dependency of those lines on density (linearly for the resonance lines and quadratic for recombination lines, such as H$\alpha$). 

The abundances of the individual elements are adjusted such that the observed strength of the spectral lines belonging to the corresponding element are reproduced best by the model.

\section{Stellar and wind parameters}
\label{sect:parameters}

The stellar and wind parameters of the investigated donor stars are listed in Table\,\ref{table:parameters} together with the corresponding error margins. For those physical quantities that are directly obtained from the spectral fit ($T_*, \log g, \log L, v_\infty, \beta, \dot{M}, E_{B-V}, R_V, v \sin i$, abundances), the corresponding errors are estimated by fixing all parameters but one and varying this parameter until the fit becomes significantly worse. For those quantities that follow from the fit parameters, the errors are estimated by linear error propagation. We do not account for uncertainties in the orbital parameters, since they are often not known. Moreover, the quoted errors do not account for systematic uncertainties, e.g., because of the simplifying assumptions of the models such as spherical symmetry. 

\begin{table*}[htbp]
    \caption{Inferred stellar and wind parameters} 
    \label{table:parameters}
    \centering  
    \renewcommand{\arraystretch}{1.3}
    \begin{tabular}{lcccccc}
        \hline\hline 
        HMXB type & 
        persistent                &
        intermediate              &
        \multicolumn{2}{c}{--------\quad SFXT \quad--------}  &
        persistent\,/ & Be X-ray   \rule[0mm]{0mm}{3.5mm} \\
        &  &  &  & & Oe X-ray & \\
        Name &   HD\,153919   &        BD+60\,73     &       LM\,Vel               &         HD\,306414      &        BD+53\,2790      &        HD\,100199  \\
        Spectral type & {\tiny O6\,Iafpe}   &   {\tiny BN0.7\,Ib}         &  {\tiny O8.5\,Ib-II(f)p}     &  {\tiny B0.5\,Ia} &  {\tiny O9.5\,Vep} & {\tiny B0\,IIIne} \\
        Alias name & {\scriptsize 4U\,1700-37 } & {\scriptsize IGR\,J00370+6122 } & {\scriptsize IGR\,J08408-4503 } & {\scriptsize IGR\,J11215-5952 } & {\scriptsize 4U\,2206+54} & {\scriptsize IGR\,J11305-6256 } \\
        \hline  
        $T_{\ast}$ (kK)                 & 
        $35^{+2}_{-3}$        & 
        $24^{+1}_{-1}$       & 
        $30^{+3}_{-3}$                  & 
        $25^{+1}_{-1}$                 & 
        $30^{+3}_{-3}$          & 
        $30^{+2}_{-3}$\rule[0mm]{0mm}{3.5mm}\\
        $T_{2/3}$ (kK)                  & 
        $34$        & 
        $23$       & 
        $29$                  & 
        $24$                 & 
        $30$          & 
        $30$        \\
        $\log g_\ast$ (cm\,s$^{-2}$)   & 
        $3.4^{+0.4}_{-0.4}$   & 
        $2.9^{+0.1}_{-0.1}$  & 
        $3.2^{+0.2}_{-0.2}$            & 
        $2.8^{+0.2}_{-0.2}$            & 
        $3.8^{+0.3}_{-0.5}$     & 
        $3.6^{+0.2}_{-0.2}$   \\
        $\log L$ ($L_\odot$)            & 
        $5.7^{+0.1}_{-0.1}$   & 
        $4.9^{+0.1}_{-0.1}$  & 
        $5.3^{+0.1}_{-0.1}$             & 
        $5.4^{+0.1}_{-0.1}$            & 
        $4.9^{+0.1}_{-0.1}$     & 
        $4.4^{+0.1}_{-0.1}$   \\
        $v_{\infty}/10^3$ (km\,s$^{-1}$)& 
        $1.9^{+0.1}_{-0.1}$   & 
        $1.1^{+0.1}_{-0.2}$  & 
        $1.9^{+0.1}_{-0.1}$             & 
        $0.8^{+0.2}_{-0.1}$            & 
        $0.4^{+0.1}_{-0.1}$     & 
        $1.5^{+0.3}_{-0.3}$   \\
        $\beta$\tablefootmark{a}         & 
        $2^{+1}_{-1}$   & 
        $1.2^{+0.6}_{-0.4}$      & 
        $1.4^{+0.4}_{-0.4}$                 & $3^{+1}_{-1}$                  & 
        $1.0$                   & 
        $0.8$                 \\
        $R_\ast$ ($R_\odot$)            & 
        $19^{+5}_{-6}$        & 
        $17^{+4}_{-4}$       & 
        $17^{+6}_{-5}$                  & 
        $28^{+6}_{-5}$                 & 
        $11^{+4}_{-4}$           & 
        $6^{+2}_{-2}$        \\
        $R_{2/3}$ ($R_\odot$)           & 
        $20$        & 
        $18$       & 
        $17$                  & 
        $31$                 & 
        $11$           & 
        $6$        \\
        $D$                             &
        $20^{+50}_{-15}$             & 
        $20^{+50}_{-16}$    & 
        $20^{+10}_{-5}$                  & 
        $20^{+10}_{-10}$                 & 
        10\tablefootmark{b}          & 
        10\tablefootmark{b}        \\
        $\log \dot{M}$ ($M_\odot \mathrm{yr}^{-1}$)& $-5.6^{+0.2}_{-0.3}$ &     
        $-7.5^{+0.1}_{-0.2}$ &    
        $-6.1^{+0.2}_{-0.2}$      & 
        $-6.5^{+0.2}_{-0.2}$                & $-7.5^{+0.3}_{-0.3}$           &   
        $-8.5^{+0.5}_{-0.5}$         \\
        $v \sin i$ (km\,s$^{-1}$)       & 
        $110^{+30}_{-50}$      & 
        $120^{+20}_{-20}$    & 
        $150^{+20}_{-20}$               & $60^{+20}_{-20}$               & 
        $200^{+50}_{-50}$       & 
        $230^{+60}_{-60}$     \\
        $M_{V,\mathrm{John}}$ (mag)     & 
        $-6.4$         & 
        $-5.3$        & 
        $-5.8$                   & 
        $-6.6$                  & 
        $-4.7$           & 
        $-3.5$         \\
        $X_{\rm H}$ (mass fr.)\tablefootmark{c}& 
        $0.65^{+0.1}_{-0.2}$ &
        $0.45^{+0.1}_{-0.1}$& 
        $0.5^{+0.1}_{-0.1}$                        & 
        $0.6^{+0.13}_{-0.2}$                        & 
        $0.7375$                & 
        $0.7375$              \\
        $X_{\rm C}/10^{-3}$ (mass fr.)\tablefootmark{c} & 
        $2.5^{+2}_{-1}$     & 
        $0.5^{+0.2}_{-0.2}$            & 
        $2.5^{+1.5}_{-1.0}$       & 
        $0.25^{+0.15}_{-0.10}$              & 
        2.37                  & 
        2.37                  \\
        $X_{\rm N}/10^{-3}$ (mass fr.)\tablefootmark{c} & 
        $2.0^{+2}_{-1}$ & 
        $2.5^{+1.5}_{-1.0}$    & 
        $2.0^{+1.0}_{-1.0}$                & 
        $4.0^{+2}_{-2}$         & 
        0.69                    & 
        0.69                  \\
        $X_{\rm O}/10^{-3}$ (mass fr.)\tablefootmark{c} & 
        $3^{+2}_{-1}$  & 
        $3^{+1}_{-1}$     & 
        $6^{+2}_{-2}$       & 
        $6^{+4.0}_{-2.5}$         & 
        5.73                    & 
        5.73                  \\
        $X_{\rm Si}/10^{-4}$ (mass fr.)\tablefootmark{c}& 
        $3^{+3}_{-2}$   & 
        $4^{+1}_{-2}$       & 
        $6^{+3}_{-3}$             & 
        $10^{+5}_{-3}$           & 
        6.65                    & 
        6.65                  \\
        $X_{\rm Mg}/10^{-4}$ (mass fr.)\tablefootmark{c}& 
        $6.92$   & 
        $9^{+3}_{-3}$      & 
        $5^{+2}_{-2}$    & 
        $5^{+4}_{-2}$       & 
        6.92                    & 
        6.92                  \\
        $E_{B-V}$ (mag)               & 
        $0.50^{+0.01}_{-0.01}$ &
        $0.85^{+0.01}_{-0.01}$&
        $0.44^{+0.01}_{-0.01}$           &
        $0.83^{+0.01}_{-0.01}$         &
        $0.595^{+0.015}_{-0.01}$ & 
        $0.34^{+0.01}_{-0.01}$\\
        $R_V$ (reddening law\tablefootmark{d})         & 
        $3.1$ (Seaton)  &
        $2.8^{+0.1}_{-0.1}$ (Cardelli)& 
        $3.1$ (Seaton)                & 
        $3.0^{+0.1}_{-0.1}$ (Cardelli) & 
        $3.1$ (Seaton)          & 
        $3.1$ (Seaton)\\
        $M_\mathrm{spec}$ ($M_\odot$)   & 
        $34^{+100}_{-28}$     & 
        $8^{+8}_{-4}$        & 
        $16^{+29}_{-11}$                & 
        $18^{+24}_{-11}$               & 
        $27^{+67}_{-23}$        & 
        $6^{+9}_{-4}$ \\
        $a_2$ ($R_\ast$)                 & 
        $1.6^{+1.5}_{-0.4}$   & 
        $2.9^{+3.2}_{-2.8}$  & 
        $2.9^{+1.6}_{-0.6}$             & 
        $12^{+5}_{-3}$                 & 
        $5.4^{+4.3}_{-1}$                 & 
        -                \\
        $v_\mathrm{orb, apa}$ (km\,s$^{-1}$)&
        $500^{+900}_{-300}$&
        $90^{+50}_{-30}$     & 
        $120^{+200}_{-60}$              & 
        $30^{+30}_{-20}$               & 
        $200^{+400}_{-200}$     & 
        -                \\
        $v_\mathrm{orb, peri}$ (km\,s$^{-1}$)&
        $500^{+900}_{-300}$&
        $300^{+200}_{-90}$  & 
        $500^{+500}_{-200}$             & 
        $300^{+200}_{-100}$            & 
        $400^{+600}_{-200}$     & 
        -                \\
        $v_\mathrm{wind, apa}$ (km\,s$^{-1}$)&
        $400^{+600}_{-300}$& 
        $850^{+50}_{-40}$   & 
        $1400^{+200}_{-200}$            & 
        $730^{+30}_{-20}$              & 
        $350^{+30}_{-30}$       & 
        -                \\
        $v_\mathrm{wind, peri}$ (km\,s$^{-1}$)&
        $400^{+600}_{-300}$&
        $200^{+200}_{-200}$ & 
        $30^{+600}_{-30}$             & 
        $220^{+200}_{-70}$             & 
        $300^{+50}_{-40}$       & 
        -                \\
        $R_\mathrm{rl, apa}\,(R_\ast)$\tablefootmark{e}&
        $1.1^{+0.5}_{-0.2}$&
        $1.5^{+0.3}_{-0.2}$&
        $1.6^{+0.6}_{-0.3}$       & 
        $2.6^{+0.7}_{-0.3}$            & 
        $1.9^{+0.9}_{-0.3}$     & 
        -                  \\
        $R_\mathrm{rl, peri}\,(R_\ast)$\tablefootmark{e}&
        $1.1^{+0.5}_{-0.2}$&
        $0.83^{+0.12}_{-0.07}$&
        $0.70^{+0.19}_{-0.08}$& 
        $1.6^{+0.3}_{-0.2}$            &
        $1.8^{+0.5}_{-0.2}$      & 
        -              \\
        \hline 
    \end{tabular}
    \tablefoot{
        \tablefoottext{a}{Values larger than unity refer to the second exponent in a double-$\beta$ law (see Sect.\,\ref{sect:models} for details).}
        \tablefoottext{b}{We were not able to determine the precise clumping factor (see Appendix.\,\ref{sec:comments} for details)}
        \tablefoottext{c}{Entries without errors are fixed to solar abundances \citep{Asplund2009} }
        \tablefoottext{d}{References: \citet{Seaton1979}; \citet{Cardelli1989}}
        \tablefoottext{e}{Calculated via the approximation presented by \citet{Sepinsky2007}, assuming the orbital parameters given in Table\,\ref{table:orbital}}
    }
\end{table*}

\subsection{Comparison with single OB-type stars}
\label{subsect:ob_comp}

The winds of massive stars are characterized by a number of quantities, such as $\dot{M}$, $v_\infty$, or $D$. Since only a low number of donor-star winds have been analyzed by means of sophisticated atmosphere models, it is statistically unfavorable to pursue comparisons for individual wind parameters. Therefore, we use the so-called modified wind momentum $D_\mathrm{mom}$ to evaluate the winds of the donor stars. The modified wind momentum is defined as
\begin{equation}
\label{eq:dmom}
D_\mathrm{mom} = \dot{M} v_\infty R_*^{1/2}~.
\end{equation} 

\begin{figure}[tbp]
    \centering
    \includegraphics[width=\hsize]{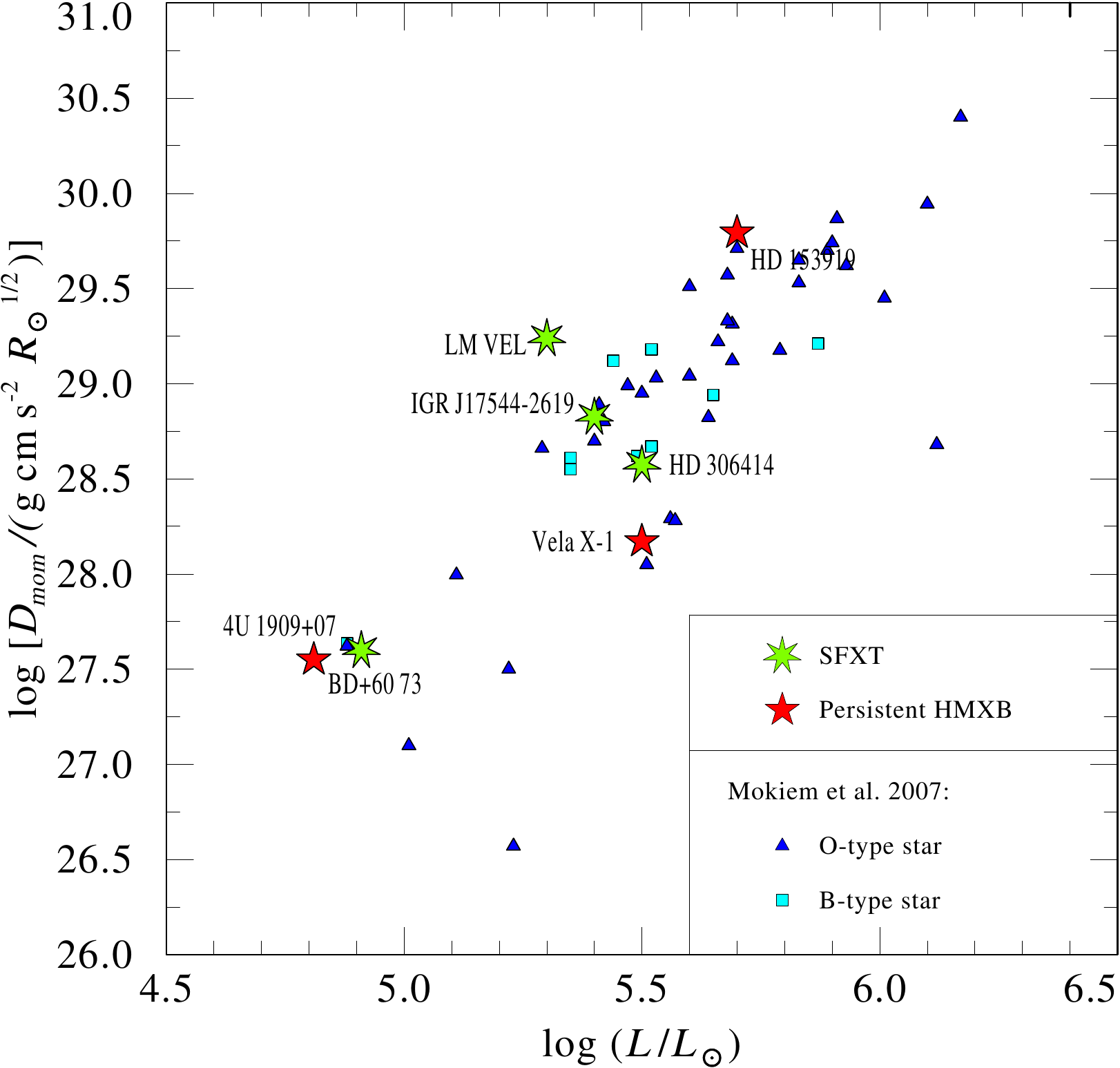}
    \caption{
        Modified wind momentum over the luminosity. The SFXTs and the persistent HMXBs are shown by green and red asterisks, respectively. In addition to the objects investigated in this work, we also show the results obtained by \citet{Martinez-Nunez2015} and \citet{Gimenez-Garcia2016}. The dark blue triangles and light blue squares depict the analyses compiled by \citet{Mokiem2007} for O and B-type stars, respectively.
    }
    \label{fig:dmom}
\end{figure}

In Fig.\,\ref{fig:dmom}, we plot $D_\mathrm{mom}$ over the luminosity. A tight linear relation between the luminosity and the modified wind momentum is predicted by the line-driven wind theory 
\citep{Kudritzki1995,Puls1996,Kudritzki1999}. 
This so-called wind-momentum luminosity relation (WLR) is observationally confirmed for a variety of massive stars \citep[e.g.,][]{Kudritzki1999,Kudritzki2002,Massey2005,Mokiem2007}. Exceptions are certain categories of objects such as the so-called weak-wind stars \citep{Bouret2003,Martins2005,Marcolino2009,Shenar2017}, where most of the wind mass-loss might be hidden from spectral analyses based on optical and UV data \citep[see e.g.,][]{Oskinova2011,Huenemoerder2012}. In addition to the stars analyzed in this work, we also plot in Fig.\,\ref{fig:dmom} the results obtained by \citet{Gimenez-Garcia2016} and \citet{Martinez-Nunez2015} for the donor stars in one SFXT (IGR\,J17544-2619) and two persistent HMXBs (Vela X-1 and U9\,1909+07) as well as the values compiled by \citet{Mokiem2007} for Galactic O and B-type stars. 

The donor stars in the investigated HMXBs fall in the same parameter regime as observed for other Galactic OB-type stars. Moreover, Fig.\,3 shows that these donors also follow the same WLRs as other massive stars in the Galaxy, indicating that the fundamental wind properties of the donor stars in wind-fed HMXBs are well within the range of what is expected and observed for these kind of massive stars. 

\subsection{Wind parameters of SFXTs versus those of HMXBs}
\label{subsect:sfxts-persistent}

A comparison between the wind parameters of the donor stars in SFXTs with those in persistent HMXBs reveals that there is no general distinction (see Table\,\ref{table:parameters}).
For example, HD\,153919 and LM\,Vel both have winds with a high terminal velocity of 1900\,$\mathrm{km}\,\mathrm{s}^{-1}$, but the former is a persistent HMXB, while the latter is a SFXT. 
Moreover, we find SFXTs with quite different wind properties: while LM\,Vel exhibits a fast stellar wind and a relatively high mass-loss rate, HD\,306414 has a significantly slower wind ($v_\infty = 800\,\mathrm{km}\,\mathrm{s}^{-1}$) and a low mass-loss rate. In fact, the parameters of HD\,306414 are quite similar to those of Vela\,X-1 \citep{Gimenez-Garcia2016}, while Vela\,X-1 is a persistent source in contrast to HD\,306414.

The wind properties are important for characterizing the donor stars. However,
the accretion onto the compact object and, consequently, the X-ray properties of a system depend on the wind conditions at the position of the compact object. Based on the orbital parameters (Table\,\ref{table:orbital}), we determine the wind velocity at the apastron and periastron positions of the NS (see Table\,\ref{table:parameters}). 
As described in Sect.\,\ref{sect:models}, we assume a double-$\beta$ law (with the second $\beta$ exponent given in Table\,\ref{table:parameters}) for the wind velocity in the supersonic regime. However, the double-$\beta$ law as well as the single $\beta$-law might not be a perfect representation of the wind structure in some HMXBs \citep{Sander2018a}.
Moreover, the wind velocity in the direction to the NS might be reduced because of the influence of the X-rays on the wind structure \citep{Krticka2015,Sander2018a}. Thus, the real wind velocity could be slightly lower than what we constrain here. However, we do not expect that this effect is significant for the objects in our sample because of the relatively low X-ray luminosities of the NSs (see Table\,\ref{table:Lxray}). 
A more detailed investigation will be presented in a forthcoming publication using hydrodynamic atmosphere models.

From Table\,\ref{table:parameters}, we clearly see that the velocities of the donor star winds at the position of the NSs ($v_\mathrm{wind, peri}$ \& $v_\mathrm{wind, apa}$) are significantly lower than the corresponding terminal wind velocities ($v_\infty$). 
We note that the value of the $\beta$ velocity-law derived in this work defines the wind structure and as such has an influence on the wind velocity determined for the position of the NS. Low $\beta$ values result in higher velocities compared to high $\beta$ values.
For the extreme case of HD\,304614, the uncertainty from the spectral fit is $\pm 1$ for the second exponent of the double-$\beta$ law. This uncertainty results in an error of less than 5\,\% for the wind velocity at the position of the NS during apastron, while it is about 30\,\% during periastron. These errors are significantly smaller than the those resulting from the orbital configuration, which are the main source for the errors quoted in Table\,\ref{table:parameters}.

The wind velocities at the position of the NSs are modulated with
the orbital configurations of the systems. An intriguing example is LM\,Vel: while the wind velocity at apastron is 1400\,$\mathrm{km}\,\mathrm{s}^{-1}$, it is as low as 30\,$\mathrm{km}\,\mathrm{s}^{-1}$ at periastron.
In contrast, the system harboring HD\,153919 (4U\,1700-37), the only truly persistent source in our sample, exhibits a negligible eccentricity and, therefore, a stable wind velocity at the position of the NS.  
This velocity is about 20\,\% of its terminal value, while the wind velocity at apastron in the SFXTs is $>70\,\%$ of $v_\infty$. 
In general, it seems that in SFXTs, the velocity of the donor star winds at the periastron position of the NSs is lower than in the persistent sources. During apastron passage this situation appears to be reversed.
Hence, we can conclude that the wind velocities at the position of the NS are significantly modulated by the orbital configuration in the SFXTs. 
This suggests that the orbits might play an important role in the dichotomy of wind-fed HMXBs as already proposed by \citet{Negueruela2006}. In general, this dichotomy likely reflects a complex interplay between the donor-star parameters, the orbital configuration, and the NS properties.

\subsection{Relative velocities and constraints on the formation of temporary accretion disks}
\label{subsect:velos}

Another interesting discovery is that the donor star wind velocity at periastron in all studied systems is within the uncertainties statistically indiscernible from the NSs orbital velocity. 
According to \citet{Wang1981} these conditions are favorable for the formation of an accretion disk around the NS.
Such a disk would act as a reservoir and might allow for X-ray outbursts peaking after periastron passage, and should also modify the X-ray light curve \citep[see e.g., ][]{Motch1991}. The formation of accretion disks regularly during periastron could potentially also influence the evolution of the spin period of the neutron star.

To check whether an accretion disk can form, we adopt the prescription from \citet{Wang1981} in the formulation given by \citet{Waters1989}. According to these studies, an accretion disk can form if 
\begin{eqnarray}
\begin{aligned}
\label{eq:disk}
v_\mathrm{rel} \leq 304\,\eta^{1/4} 
\left( \frac{P_\mathrm{orb}}{10\,\mathrm{d}}    \right)^{-1/4} 
\left( \frac{M_\mathrm{NS}}{M_\odot}             \right)^{5/14} 
\left( \frac{R_\mathrm{NS}}{10^6\,\mathrm{cm}}   \right)^{-5/28}\\
\left( \frac{B_0}{10^{12}\,\mathrm{G}} \right)^{-1/14} 
\left( \frac{L_\mathrm{X}}{10^{36}\,\mathrm{erg/s}} \right)^{1/28}\,\mathrm{km}\,\mathrm{s}^{-1}\,,
\end{aligned}
\end{eqnarray} 
where $P_\mathrm{orb}$ is the orbital period in days and $\eta$ describes the efficiency of the angular momentum capture. The NS properties enter with the magnetic flux density $B_0$, the X-ray luminosity $L_\mathrm{X}$, the NS mass $M_\mathrm{NS}$, and radius $R_\mathrm{NS}$. 

With the help of Eq.\,(\ref{eq:disk}), we can thus estimate whether an accretion disk around the NSs in our target systems would form. For $R_\mathrm{NS}$, we assume $1.1\times10^6\,\mathrm{cm}$ based on the estimates by \citet{Oezel2016}. 
We assume a magnetic flux density of $B_0 = 10^{12}\,\mathrm{G}$ for all NSs in our sample. The only exception is BD+53\,2790, where \citet{Torrejon2018} constrain the magnetic field of the NS to $B_0 > 2 \times 10^{13}\,\mathrm{G}$. We also assume the canonical NS mass  $M_\mathrm{NS} = 1.4\,M_\odot$ \citep{Thorsett1999}. The only exception is the NS companion of HD\,153919, for which \citet{Falanga2015} derive a mass of $1.96\,M_\odot$. Moreover, we set the efficiency factor to $\eta = 1$, as expected in the presence of an accretion disk \citep{Waters1989}. Based on these assumptions, 
we find that no disks are predicted to form in any of our target systems. 
Note that Eq.\,(\ref{eq:disk}) is strictly valid only for circular systems. Moreover, if the X-ray luminosity of the SFXTs is higher during an outburst than during our \textsl{Swift} observations, we might obtain a different result. However, even for $Lx = 10^{38}\,\mathrm{erg}\,\mathrm{s}^{-1}$ no accretion disks are predicted to form.

Recent detailed studies of wind dynamics in the vicinity of an accreting NS have been performed by \citet{ElMellah2019b}. Their 3-D simulations show that when orbital effects are dynamically important, the wind dramatically departs from a radial outflow in the NS vicinity and the net angular momentum of the accreted flow could be sufficient to form a persistent disk-like structure. On the other hand, the 3-D hydrodynamic models by \citet{Xu2019} show that in flows that are  prone to instability, such as stellar winds, the disks are not likely to form. In support of this, observations do not indicate presence of stable accretion disks in HMXBs with NS components \citep[e.g.,][]{Bozzo2008}. Thus, the question of persistent disk formation remains open. Our spectral models, which rely on spherical symmetry, are capable of reproducing the line shapes formed in the stellar wind (e.g., lines with P Cygni profiles); this argues in the favor of the models where the wind flow is strongly bent only in a limited volume close to the NS.     

We highlight that the orbital velocity cannot be neglected, since it is comparable to the wind velocity at the position of the NS. Thus, it needs to be accounted for when estimating the mass accretion rate from the donor-star wind according to the Bondi-Hoyle-Lyttleton mechanism. Consequently, the orbital velocity is important for predicting the X-ray luminosity (see also Sect.\,\ref{sect:intro} and \ref{sect:x-rays}).

\subsection{Abundances}
\label{subsect:abundances}

In Table\,\ref{table:parameters}, we also list the chemical abundances for our program stars. 
Abundances that are derived from the spectral fits are given with the estimated errors. For those elements where only insufficient diagnostics are available, the abundances are fixed to the solar values, and the corresponding entries in Table\,\ref{table:parameters} are given without errors. 

Two thirds of our sample (HD\,153919, BD+60\,73, HD\,306414, and LM\,Vel) shows a significant depletion of hydrogen compared to the primordial abundance. For BD+53\,2790 and HD\,100199 no deviation from this value could be detected. 
Nitrogen is enriched with respect to the solar value \citep{Asplund2009} in all investigated wind-fed HMXBs. 
HD\,153919 and LM\,Vel exhibit a carbon abundance that is approximately solar, while carbon is subsolar in all other objects. 
The same applies to oxygen, which is depleted in all investigated objects with the exception of HD\,306414 and LM\,Vel, which shows an oxygen abundance of about $X_\mathrm{O} = X_{\mathrm{O}, \odot}$ and $1.1\,X_{\mathrm{O}, \odot}$, respectively.

\citet{Crowther2006c} determine CNO abundances for 25 Galactic OB-type supergiants. They constrain mean [N/C], [N/O], and [C/O] logarithmic number ratios (relative to solar) of +1.10, +0.79, and -0.31, respectively. Only BD+60\,73 appears to be fully consistent with these results, while the other objects in our sample exhibit conspicuous abundance patterns. The [C/O] ratio of HD\,153919 and LM\,Vel (0.31 and 0.01) is significantly higher than the average values derived by \citet{Crowther2006c}, while it is substantially lower for HD\,306414 ([C/O] = -1.0). 

In general, silicon and magnesium seems to be depleted in our program stars, with the exception of HD\,306414 and BD+60\,73. The former shows a supersolar silicon abundance, while the latter exhibits a slightly supersolar magnesium abundance. 
However, we note that the uncertainties for these abundance measurements are quite high. Hence, the results have to be interpreted with caution. 
In the next section, we will discuss these abundance patterns in an evolutionary context.

\section{Stellar evolutionary status}
\label{sect:evolution}

The detailed investigation of HMXBs offers the possibility to constrain open questions of massive star evolution, SN kicks, and common envelope (CE) phases.

\subsection{Common envelope evolution and NS natal kicks}
\label{subsect:evo_cce}

The formation of a HMXB is a complex process. In the standard scenario, a massive binary system initiates RLOF from the primary to the secondary. This mass transfer becomes dynamically unstable, if the secondary cannot accrete all of the material. This
often results in a CE phase that either leads to a merger or to the ejection of the primary's envelope, entailing a significant shrinkage of the binary orbit \citep[e.g.,][and references therein]{Paczynski1967,Taam2000,Taam2010,Ivanova2013}.
In the latter case, the stripped primary will undergo a core collapse forming a compact object, which can accrete matter from the rejuvenated secondary. These systems then emerge as HMXB. 

If the mass transfer is stable, or in the case of large initial orbital separations, a CE phase can be avoided.
To form a HMXB, however, this evolutionary path requires fortuitous SN kicks to reduce the orbital separation to the small values observed for the majority of these systems \citep{Walter2015}.

With the exception of HD\,306414, all investigated wind-fed HMXBs have tight orbits with periods of less than 16\,d and semi-major axes of less than $64\,R_\odot$. These separations are significantly smaller than the maximum extension of the NS progenitor. Therefore, each of these systems could indeed have already passed through a CE phase. Alternatively, the core collapse that leads to the formation of the NS was asymmetric and imparted a natal kick on the new-born NS. This reduced the orbital separations and hardened the system. 
A third possibility, in principle, is that the binary was in a close configuration from the beginning, and this has not changed because the components evolved quasi-homogeneously \citep[e.g.,][]{Maeder1987,Langer1992,Heger2000,Yoon2005,Woosley2006}.
This prevents a significant expansion of the stars, so that the system never entered a CE phase. 
However, there is no reason to suspect quasi homogeneous evolution (QHE) in our studied donor stars. 

No significant eccentricity is expected for post CE systems, which is in strong contrast to most HMXBs in our sample. Yet, the current eccentricity of these systems might be a result of the core-collapse event, suggesting that relatively large natal kicks are associated with the formation of NSs. 
This appears to be consistent with the results presented by
\citet{Tauris2017}, who find evidence that the kicks of the first SN in binaries evolving towards double neutron stars (DNSs) are on average larger than those of the second SN.
In our sample, only HD\,153919 does not show any substantial eccentricity. This is either a result of tidal circularization after the first SN, which appears plausible considering the advanced evolutionary status of HD\,153919, or of a CE phase, which however implies that the SN kick was negligible in this case. 
Although the presence of a NS in this system is strongly favored \citep{Martinez-Chicharro2018}, a BH cannot be excluded. Thus, a third possibility exists for HD\,153919. Since the formation of a BH is not necessary associated with a SN and a corresponding kick, the virtual circular orbit of HD\,153919 might be a result of a CE phase.

\subsection{Abundance pattern}
\label{subsect:evo_abund}

The atmospheric abundance pattern of evolved massive stars, such as OB-type supergiants, is often affected by CNO burning products. Those are mixed to the surface due to processes, such as rotational induced mixing \citep{Heger2000}. Accordingly, it is expected that the oxygen and carbon abundance decrease in favor of the nitrogen abundance in the course of the evolution. 
As stated in Sect.\,\ref{sect:parameters}, most of our program stars are not compatible with this scenario. 
On the one hand, HD\,153919, LM\,Vel, HD\,306414, and BD+60\,73 show hydrogen depletion and nitrogen enrichment, which point to an advanced evolution state. On the other hand, HD\,153919 and LM\,Vel have about solar carbon abundance, and HD\,306414 has a supersolar oxygen abundance.
Only for BD+60\,73 the hydrogen and the CNO abundances are consistent with an advanced evolution state according to single-star evolution. 

For HD\,153919, \citet{Clark2002} point out that its carbon overabundance has to be a result of  accretion from the NS progenitor during its Wolf-Rayet (WR) stage, more precisely during the carbon sequence WR (WC) phase. This can also serve as an explanation for the high carbon and oxygen abundances of LM\,Vel and HD\,306414, respectively. In the latter case, the NS progenitor had to reach the oxygen rich WR (WO) phase before exploding as SN. For this scenario to work, the masses of the corresponding WC and WO stars had to be below a certain limit to form NSs at the end. \citet{Woosley2019} estimate that most stars with final masses up to $6\,M_\odot$, corresponding to $9\,M_\odot$ helium core masses or $30\,M_\odot$ on the ZAMS, will leave neutron star remnants. This constraint is compatible with a few Galactic WC stars \citep{Sander2019}. If this scenario is true, it proves that the low mass WC/WO stars indeed explode as Type Ibc\,SN, instead of directly collapsing to a BH. 

Alternatively, the high carbon and oxygen abundances might be explained by pollution of material ejected during the SN explosion. 
In this case, significant enrichment by other elements such as silicon and magnesium is expected as well, based on calculation of nucleosynthesis yields from core-collapse SNe \citep[e.g.,][]{Rauscher2002,Nomoto2006,Woosley2007,Nomoto2013}.
This is in contradiction to what is derived in our spectral analyses such as the low silicon abundances in the atmospheres of HD\,153919 as well as the slightly subsolar magnesium abundance in LM\,Vel and HD\,306414. 
However, we note the supersolar magnesium abundance of HD\,306414 and BD+60\,73.

\subsection{Angular momentum transfer and projected rotational velocities}
\label{subsect:evo_vrot}

Interacting binary stars do not only exchange mass but also angular momentum. Mass transfer due to RLOF often spins up the accreting star until this mass gainer rotates nearly critical \citep{Packet1981,deMink2013}. As mentioned earlier, the orbital parameters of all objects in our sample suggest that mass transfer has occurred in these systems in the past. However, in subsequent phases (especially the presented HMXB stage) the remaining OB-type star could loose angular momentum by its wind. It is therefore interesting to check if the donor stars exhibit rapid rotation. 

We derive projected rotational velocities in the range from 60 to 230\,$\mathrm{km}\,\mathrm{s}^{-1}$. 
Interestingly, the smallest $v \sin i$ is found for HD\,306414, which might has avoided strong binary interactions in the past. 
The two OBe stars in our sample exhibit the larges projected rotational velocities (200\,$\mathrm{km}\,\mathrm{s}^{-1}$ and 230\,$\mathrm{km}\,\mathrm{s}^{-1}$). Nonetheless, we can rule out very rapid rotation for all donor stars in our sample. Using a rough approximation for the critical velocity $v_\mathrm{crit} = \sqrt{G M_\ast R_\ast^{-1}}$ (neglecting for example effects due to oblateness) and adopting the mean statistical inclination of $57\degr$, all donor stars are found to rotate far below critical. 

The $v \sin i$ distribution of Galactic OB-type stars has been investigated in many studies \citep[e.g.,][]{Dufton2006,Fraser2010,Braganca2012,Simon-Diaz2010,Simon-Diaz2014,Garmany2015}. These studies often find evidence of a bimodal distribution, showing a low $v \sin i$ peak and a group of fast rotators that extends to very high $v \sin i$ \citep[e.g.,][]{Ramirez-Agudelo2013,Simon-Diaz2014,Garmany2015}.
A similar result is obtained by \citet{Ramachandran2018} for $> 200$ OB-type stars in the Large Magellanic Cloud (LMC).
\citet{deMink2013} predict that the high $v \sin i$ peak predominately results from massive stars that were spun up because of binary interactions, while the low-velocity peak consists of single stars and binary systems that have not interacted yet.  

Based on a study of about 200 northern Galactic OB-type stars, which also accounts for the effects of macroturbulence and microturbulence, \citet{Simon-Diaz2014} find that the $v \sin i$ distribution for O and B-type supergiants peaks at 70\,$\mathrm{km}\,\mathrm{s}^{-1}$ and 50\,$\mathrm{km}\,\mathrm{s}^{-1}$, respectively. 
Comparing this with the projected rotational velocities of our sample, it appears that our program stars rotate on average more rapidly than single OB-type stars. This is in accordance with mass accretion in the past. The only exception might be HD\,306414.

For the O-type components in six Galactic WR\,+\,O binaries, \citet{Shara2017} derive  rotational velocities. Those are expected to be nearly critical, since the O-type stars are spun up by RLOF from the WR progenitor. However, \citet{Shara2017} find that these stars spin with a mean rotational velocity of $350\,\mathrm{km}\,s^{-1}$, which is about 65\,\% of their critical value. They argue that a significant spin-down even on the short timescales of the WR-phase (a few hundred thousand years) must have taken place. The rotational velocities derived for our donor stars are substantially lower than those of the O-type components in the WR binaries. Compared to these objects, the evolution time scales of our stars are significantly larger (a few million years). Thus, our donor stars might had more time to spin down, which would be consistent with their lower rotational velocities. In this picture, our results and those by \citet{Shara2017} coincide nicely.

\subsection{Mass-luminosity relation}
\label{subsect:evo_mass}

In binary systems it is expected that the mass gainer is internally mixed because of angular momentum transfer. Therefore, the mass gainer should be overluminous compared to single stars of the same mass \citep[e.g.,][]{Vanbeveren1994}. 
To investigate whether this is the case for our program stars, we compare the spectroscopic masses constrained in this work with masses from stellar-evolution tracks. The latter are obtained with the BONNSAI Bayesian statistics tool \citep{Schneider2014}. 
Using stellar and wind parameters ($T_\ast, \log L, \log g, v \sin i, X_\mathrm{H}, \dot{M}$) and their corresponding errors as input, the BONNSAI tool interpolates between evolutionary tracks calculated by \citet{Brott2011}. Based on this set of single star evolution tracks, the tool predicts the current mass that an object with these parameters would have, if it has evolved like a single star. The correlated parameters are listed and compared in Table\,\ref{table:bonnsai}.

For BD+60\,73 and HD\,306414, evolution masses could not be derived in this way, since the parameters of these stars are not reproduced by any of the underlying stellar evolution models. BD+53\,2790 exhibits a spectroscopic mass that is 35\,\% larger than its evolution mass. HD\,153919 and LM\,Vel seem to be overluminous for their current mass.

\subsection{Future evolution}
\label{subsect:evo_future}

Unfortunately, binary evolution tracks that would be applicable to the HMXBs investigated in this work are not available. 
Nevertheless, the future evolution of our targets can be discussed based on their current orbital configuration and the stellar properties of the donor stars. All investigated systems are compact enough that the donor stars, in the course of their further evolution, will expand sufficiently to eventually fill their Roche lobe, initiating direct mass transfer to their NS companions. Whether or not this mass transfer is stable will significantly influence the further evolution and the final fate of these HMXBs.

The stability of the mass-transfer in such systems has recently received increased attention. 
\citet{vandenHeuvel2017} study whether this mass-transfer would lead to a (second) CE phase and whether this would  result in a merger. They conclude that the mass-transfer is indeed unstable for a broad parameter range, and that the vast majority of the known HMXBs, consisting of supergiants with NS companions (>95\,\%) would not survive the spiral-in within a CE phase. 
Applying their findings to our results, and assuming a NS mass of $1.4\,M_\odot$, suggests that also all systems investigated in this work will enter a CE phase that leads to a merger. 
The same can be concluded from a comparison of the stellar and orbital parameters of our HMXBs with the CE-ejection solutions calculated by \citet{Kruckow2016} for massive binary systems . For all our objects, the minimal orbital separation is significantly lower than $100\,R_\odot$, while the spectroscopic masses are higher than $8\,M_\odot$. Comparing these constraints with the solutions presented by \citet[][see their figure\,2]{Kruckow2016} suggests that the systems studied in this work are not able to eject the CE in the upcoming CE phase.
These findings are consistent with conclusions by previous studies \citep[e.g.,][]{Podsiadlowski1994,vandenHeuvel2017,Tauris2017}.

If the systems studied in this work merge, they will form so-called Thorne–Żytkow objects \citep[TŻO, ][]{Thorne1975,Thorne1977}. \citet{Cannon1992} already discuss HMXBs as a potential source of TŻOs, identifying this as one of two possible channels. \citet{Podsiadlowski1995} estimate the number of TŻOs in the Galaxy to be 20-200. Thorne–Żytkow objects will likely appear as red supergiants (RSGs) \citep{Biehle1991,Cannon1993}, which are only distinguishable from normal RSGs by means of specific abundance patterns. These abundances are a result of the extremely hot non-equilibrium burning processes, that allow for interrupted rapid proton addition \citep{Thorne1977,Cannon1993}. The first promising candidate for a TŻO is identified by \citet{Levesque2014}.
According to \citet{Tauris2017}, a few to ten percent of the luminous red supergiants ($L \ge 10^{5}\,L_\odot$) in the Galaxy are expected to harbor a NS in their core. 

Alternatively, TŻOs might appear as WN8 stars. This is suggested by \citet{Foellmi2002} because of the peculiar properties of these class of objects, such as the low binary fraction, strong variability, and the high percentage of runaways. Recently, this has been proposed to be a valid scenario for WR\,124 \citep{Toala2018}. Based on population synthesis models, already \citet{deDonder1997} have proposed that WR stars with compact objects at their center should exist. They denote these objects as ``weird'' WR stars.
In view of the above results, we are inclined to conclude that the binaries examined in this work will presumably form some kind of TŻOs in the future.

However, a certain fraction of the HMXB population obviously survives, since we see compact DNS systems. 
If the HMXBs can avoid a merger in the imminent CE phase, they 
will likely undergo an additional phase of mass transfer according to the Case BB scenario \citep{Tauris2015,Tauris2017}. This will lead to an ultra stripped star, which will explode as a Type Ib/Ic SN, leaving a NS. Since the associated kick will likely be small \citep{Tauris2017}, the binary system will presumably stay intact, forming a DNS. 

Independent of the future evolution of the HMXBs investigated in this study, we highlight that HMXBs and their properties offer the possibility to falsify stellar evolution scenarios and population synthesis models predicting event rates of double degenerate mergers. 
These simulations often also include some kind of HMXB evolution phase. Thus, the properties of the HMXB population can be used to constrain these models.
Therefore, further studies analyzing a large fraction of the HMXB population are imperative.

\section{Efficiency of the accretion mechanism}
\label{sect:x-rays}

The X-ray luminosity $L_\mathrm{X}$ of the accreting NS in our HMXBs is related to the accretion rate $S_\mathrm{accr}$ via the accretion efficiency parameter $\epsilon$: 
\begin{equation}
\label{eq:lx2}
L_\mathrm{X} = \epsilon S_\mathrm{accr} c^{2}\,.
\end{equation} 
The actual value of the accretion efficiency depends on the detailed physics of the accretion mechanism. 
Comparing the X-ray luminosities measured with \textsl{Swift} during our \textsl{HST} observation (see Table\,\ref{table:Lxray}) with theoretical expectations, we are able to put observational constraints on $\epsilon$ in some of the systems in our sample.

In the Bondi-Hoyle-Lyttleton formalism \citep[e.g.,][]{Davidson1973,Martinez-Nunez2017}, the stellar wind accretion rate, $S_\mathrm{accr}$, can be estimated as 
\begin{eqnarray}
\begin{aligned}
\label{eq:accr}
S_\mathrm{accr} \approx 1.5 \times 10^{7} \left( \frac{\dot{M}}{M_\odot\,\mathrm{yr}^{-1}} \right) \left( \frac{v_\mathrm{rel}}{10^{8} \mathrm{cm}\,\mathrm{s}^{-1}} \right)^{-4} \\  \left( \frac{M_\mathrm{NS}}{M_\odot} \right) \left( \frac{d_\mathrm{NS}}{R_\odot} \right)^{-2} S_\mathrm{Edd}\,,
\end{aligned}
\end{eqnarray}
where $d_\mathrm{NS}$ is the orbital separation and $S_\mathrm{Edd}$ is the Eddington accretion rate, which is defined as 
\begin{equation}
\label{eq:Sedd}
S_\mathrm{Edd} = \frac{L_\mathrm{Edd}}{c^{2}}\,,
\end{equation} 
with $L_\mathrm{Edd}$ being the Eddington luminosity. 
For a fully ionized plasma that only consists of helium and hydrogen, $L_\mathrm{Edd}$ can be approximated as 
\begin{equation}
\label{eq:Ledd}
L_\mathrm{Edd} \approx 2.55 \times 10^{38} \frac{M_\mathrm{NS}/M_\odot}{1 + X_\mathrm{H}}\,\mathrm{erg}\,\mathrm{s}^{-1}\,.
\end{equation}

The hydrogen mass fraction $X_\mathrm{H}$ of the accreted material is obtained from our spectral analyses. The orbital separation between the donor star and the NS as well as the relative velocity of the matter passing by the NS are phase dependent. To allow for a meaningful comparison between the observed and the predicted X-ray flux, these parameters need to be calculated for the specific phase of our simultaneous \textsl{HST} and \textsl{Swift} observation. 
This is possible for only two systems in our sample (BD+60\,73 and LM\,Vel) because an estimate of the inclination $i$ is prerequisite for these calculations.
To derive $i$, we make use of the mass function 
\begin{equation}
\label{eq:massfunc}
f(M) = \frac{M_\mathrm{NS}^{3} \sin^{3} i}{(M_\mathrm{spec} + M_\mathrm{NS})^{2}}\,.
\end{equation} 
For BD+60\,73 a mass function of $f(M) = 0.0069\,M_\odot$ is derived by \citet{Gonzalez-Galan2014}, while \citet{Gamen2015} determine $f(M) = 0.004\,M_\odot$ for LM\,Vel. 
Using the spectroscopic mass of the donor stars as derived from our spectral analyses, we are able to estimate the inclination to about $38\degr$ and $50\degr$ for BD+60\,73 and LM\,Vel, respectively. 

With the inclination at hand and the orbital period as well as the eccentricity from Table\,\ref{table:orbital}, we solve the Kepler equation numerically. This allows us to derive the phase dependent distance between the NS and the donor star $d_\mathrm{NS}$. The wind velocity at the position of the NSs during our \textsl{HST} and \textsl{Swift} observations can then be derived from the atmosphere models.

With these properties, we are able to derive the accretion efficiencies using Eq.\,(\ref{eq:lx2}) to (\ref{eq:Ledd}). 
For BD+60\,73, we obtain $\epsilon = 1.1 \times 10^{-3}$, while it is approximately a factor of two higher for LM\,Vel ($\epsilon = 2.1 \times 10^{-3}$). 
Although all stellar and wind parameters of the donor stars are constrained well, these results must be treated with some caution because of the discrepancies of the spectral fits described in Appendix\,\ref{sec:comments}.
\citet{Shakura2014} suggest that at low-luminosity states, SFXTs can be at 
the stage of quasi-spherical settling accretion when the accretion rate on to 
the NS is suppressed by a factor of $\sim 30$ relative to the Bondi-Hoyle-Lyttleton 
value. This might be sufficient to explain the low accretion efficiency 
deduced for LM\,Vel and BD+60\,73. Alternatively, \citet{Grebenev2007} and \citet{Bozzo2008} 
suggest that a magnetic gating or a propeller mechanism could strongly inhibit 
the accretion in SFXTs.

\section{Wind accretion vs. Roche-lobe overflow}
\label{sect:roche}

All HMXBs in our sample are thought to accrete matter only from the donor star wind or from its decretion disk. This perception is called into question by our analyses. For a subset of our sample RLOF during periastron passage seems plausible. 

The Roche-lobe radii of the donor stars in our sample are estimated using a generalization of the fitting formula by \citet{Eggleton1983} for nonsynchronous, eccentric binary systems provided by \citet{Sepinsky2007}. For BD+60\,73 and LM\,Vel, the Roche-lobe radius at periastron, $R_\mathrm{rl, peri}$, is smaller than the stellar radius (see  in Table\,\ref{table:parameters}). During this orbital phase, matter can be directly transferred to the NS via the inner Lagrangian point. We note that this finding has no influence on the estimates performed in the previous section since the \textsl{HST} and corresponding \textsl{Swift} observations of these sources were performed during a quiescent X-ray phase.

Interestingly, both of these sources are classified as X-ray transients. 
For BD+60\,73, the X-ray light curve folded with the orbital phase peaks around $\phi \approx 0.2$, corresponding to 3 to 4\,d after periastron \citep{Gonzalez-Galan2014}. This behavior is usually attributed to an increased wind accretion-rate during periastron passage because of the lower wind velocity and higher wind density during this phase. However, for BD+60\,73, the reason could be direct overflow of matter, which follows the gravitational potential. The delay between periastron passage and outburst might be due to inhibition of direct accretion onto the NS because of magnetic and centrifugal gating mechanisms \citep{Illarionov1975,Grebenev2007,Bozzo2008}.

For LM\,Vel, the X-ray outbursts cluster around periastron as well \citep{Gamen2015}. In contrast to BD+60\,73, however, the outbursts are also observed prior to periastron passage ($\phi = 0.84-0.07$), suggesting that in this case a combination of donor-wind capture and RLOF might feed the accretion.

For these two systems, the amount of mass transfer via RLOF needs to be relatively limited, since otherwise these systems are expected to quickly enter a CE phase. Moreover, we note that our estimates of the Roche-lobe radius should be treated with caution, since some of the orbital parameters of our binary systems, such as the inclination, are not well constrained. 

Hydrodynamical simulations \citep{Mohamed2007} suggest that a further mode of mass transfer plays a role in certain binary systems.
This so-called wind Roche-lobe overflow (WRLOF) invokes a focusing of the primary stellar-wind towards the secondary. Recently, \citet{ElMellah2019a} have suggested that this mechanism is chiefly responsible for the formation of so-called ultra-luminous X-ray sources (ULXs), and that it also plays a role in certain HMXBs. Wind Roche-lobe overflow gets important when the radius were the wind is accelerated beyond the escape velocity is comparable to the Roche-lobe radius \citep{Mohamed2007,Abate2013}. This condition is fulfilled for all wind-fed HMXBs in our sample (HD\,153919, HD\,306414, BD+60\,73, and LM\,Vel). However, the detailed calculations by \citet{ElMellah2019a} suggest that this might be a too crude criterion. Their scenario for NSs is roughly applicable to HD\,306414. For this object, their model predicts WRLOF for periastron, but not for apastron.  

WRLOF seems to be a possible mass-transfer mechanism in wind-fed HMXBs, but presumably not during all orbital phases. 
Mass-transfer in these systems can be significantly higher compared to the classical Bondi-Hoyle-Lyttleton mechanism \citep{Podsiadlowski2007}. However, this is not directly reflected in the X-ray luminosities of these objects, which are moderate (see e.g., Table\,\ref{table:Lxray}). So, 
an effective gating mechanism seems to be at work in these systems that hampers the accretion of the transferred material (see also discussion in \citealt{ElMellah2019a} on Vela\,X-1).

\section{Summary and Conclusions}
\label{sect:conclusions}

For this study, we observed six HMXBs with the \textsl{HST} STIS and secured high S/N, high resolution UV spectra. Simultaneously to these \textsl{HST} observations, we obtained \textsl{Swift} X-ray data to characterize the X-ray emission of the NSs. These data sets were used to determine the wind and stellar parameters of the donor stars in these HMXBs by means of state of the art model atmospheres, accounting for the influence of the X-rays on the donor-star atmosphere. The wind parameters of these objects were deduced for the first time. Based on these analyses, we draw the following conclusions:

\smallskip\noindent
1) The donor stars occupy the same parameter space as the putatively single OB-type stars from the Galaxy. Thus, the winds of these stars do not appear to be peculiar, in contrast to earlier suggestions.

\smallskip\noindent
2) There is no systematic difference between the wind parameters of the donor stars in SFXTs compared to persistent HMXBs. 

\smallskip\noindent
3) All SFXTs in our sample are characterized by high orbital eccentricities. Thus, the wind velocities at the position of the NS and, consequently, the accretion rates are strongly phase dependent. This leads us to conclude that the orbital eccentricity is decisive for the distinction between SFXTs and persistent HMXBs. 

\smallskip\noindent
4) In all investigated systems, the orbital velocities of the NSs are comparable to the wind velocity at their position. Therefore, the orbital velocity is important and can not be neglected in modeling the accretion or in estimating the accretion rate. 

\smallskip\noindent
5) 
Since all systems in our study have very tight orbits, the donor-star wind has not yet reached its terminal velocity when passing the position of the NS. While this has been reported earlier, it is in strong contrast to what is often implicitly assumed in the wider literature.

\smallskip\noindent
6) 
For BD+60\,73 and LM\,Vel, RLOF potentially occurs during periastron passage. Moreover, WRLOF seems plausible in a variety of HMXBs. 

\smallskip\noindent
7) 
The donor stars of HD\,153919, BD+60\,73, and LM\,Vel are in advanced evolutionary stages, as indicated by their abundance patterns. They are on the way to become red supergiants and will thus engulf their NS companion soon. 

\smallskip\noindent
8) The carbon and oxygen abundances of HD\,153919, LM\,Vel, and HD\,306414 suggest that their atmospheres were polluted by material accreted from the wind of the NS progenitor or SN ejecta.

\smallskip\noindent
9) The donor star of HD\,153919 and LM\,Vel are overluminous for their current mass.

\smallskip\noindent
10) Statistically, the donor stars in our sample rotate faster than single OB-type stars typically do, suggesting mass accretion because of RLOF in the past. This is consistent with the orbital parameters of these systems. 

\smallskip\noindent
11) Most likely, the donor stars and the NSs of the HMXBs studied in this work will merge in an upcoming CE phase, forming some kind of Thorne–Żytkow objects.

\smallskip\noindent
12) The accretion efficiency parameters $\epsilon$ of the NS in our sample are quite low, suggesting that either spherical settling accretion or a gated accretion mechanism was at work during our observations.

\begin{acknowledgements}

We thank the anonymous referee for their constructive comments.
The first author of this work (R.\,H.) is supported by the Deutsche Forschungsgemeinschaft
(DFG) under grant HA 1455/28-1.
L.\,M.\,O. acknowledges support from the Verbundforschung grant 50 OR 1809.
J.M.T. acknowledges the research grant ESP2017-85691-P.
A.\,A.\,C.\,S. is supported by the Deutsche Forschungsgemeinschaft (DFG) under grant HA 
1455/26. 
F.\,F., K.\,S., and A.\,B. are grateful for support from STScI Grant HST-GO-13703.002-A.
T.\,S. acknowledges support from the European Research Council (ERC) under the European Union's DLV-772225-MULTIPLES Horizon 2020 research and innovation programme.  
Some/all of the data presented in this paper were obtained from the Mikulski Archive for Space Telescopes (MAST). STScI is operated by the Association of Universities for Research in Astronomy, Inc., under NASA contract NAS5-26555. Support for MAST for non-\textsl{HST} data is provided by the NASA Office of Space Science via grant NNX09AF08G and by other grants and contracts.
This research has made use of the VizieR catalogue access tool, Strasbourg, France. The original description of the VizieR service was published in A\&AS 143, 23.

\end{acknowledgements}

\bibliographystyle{aa}
\bibliography{paper}


\label{onlinematerial}

\begin{appendix} 
\section{Additional tables}
\label{sec:addtables}

\begin{table*}[htbp]
    \caption{Spectroscopic data} 
    \label{table:spectra}
    \small
    \centering  
    \begin{tabular}{lllccSS}
        \hline\hline 
        Identifier &
        Wavelength &
        Instrument &
        Resolving power &
        Observation date &
        \multicolumn{1}{c}{MJD} &
        \multicolumn{1}{c}{Phase} \rule[0mm]{0mm}{3.5mm} \\
         &
        \multicolumn{1}{c}{(\AA)} &
         &
         &
         &
        \multicolumn{1}{c}{(d)} & \\
        \hline  
        HD\,153919  &  905-1187 & FUSE        & 20000   & 2003-07-30 & 52850.91896991 & 0.945  \rule[0mm]{0mm}{4.0mm}\\
        &  905-1187 & FUSE        & 20000   & 2003-07-31 & 52851.76329861 & 0.193\\
        &  905-1187 & FUSE        & 20000   & 2003-04-07 & 52736.60747685 & 0.439\\
        &  905-1187 & FUSE        & 20000   & 2003-08-02 & 52853.3525463  & 0.658\\
        & 1150-1700 & STIS/\textsl{HST}    & 45800   & 2015-02-22 & 57075.26050776 & 0.138\\
        & 3630-7170 &FEROS/ESO-2.2m&48000   & 2005-06-25 & 53546.31882176 & 0.773\\
        & 3630-7170 &FEROS/ESO-2.2m&48000   & 2009-05-03 & 54954.27310296 & 0.457\\
        & 3630-7170 &FEROS/ESO-2.2m&48000   & 2011-05-18 & 55699.24895024 & 0.816\\
        
        BD+60\,73   & 1150-1700 & STIS/\textsl{HST}    & 45800   & 2015-01-01 & 57023.49188341 & 0.841\\
        & 3630-7170 & FIES/NOT    & 25000   & 2013-01-29 & 56321.83944792 & 0.038\\
        LM\,Vel &  905-1187 & FUSE        & 20000   & 1999-12-26 & 51538.83758102 & 0.366\\
        & 1150-1700 & STIS/\textsl{HST}    & 45800   & 2015-07-16 & 57219.38183042 & 0.855\\
        & 1150-1980 & SWP/IUE     & 10000   & 1994-12-09 & 49607.75356481 & 0.709\\
        & 1150-1980 & SWP/IUE     & 10000   & 1994-12-09 & 49607.88165509 & 0.695\\
        & 1150-1980 & SWP/IUE     & 10000   & 1994-12-09 & 49607.99155093 & 0.684\\
        & 1850-3350 & LWP/IUE     & 15000   & 1994-12-09 & 49607.84041667 & 0.7  \\
        & 1850-3350 & LWP/IUE     & 15000   & 1994-12-09 & 49607.95228009 & 0.688\\
        & 3630-7170 &FEROS/ESO-2.2m&48000   & 2006-01-04 & 53739.20960784 & 0.806\\
        & 3630-7170 &FEROS/ESO-2.2m&48000   & 2007-04-18 & 54208.99984617 & 0.58 \\
        HD\,306414  & 1150-1700 & STIS/\textsl{HST}    & 45800   & 2015-08-16 & 57250.78585894 & \multicolumn{1}{c}{-} \\
        & 3630-7170 &FEROS/ESO-2.2m&48000   & 2007-01-17 & 54117.14642524 & \multicolumn{1}{c}{-} \\
        & 3630-7170 &FEROS/ESO-2.2m&48000   & 2007-02-13 & 54144.06818366 & \multicolumn{1}{c}{-} \\
        BD+53\,2790 & 1150-1700 & STIS/\textsl{HST}    & 45800   & 2015-08-15 & 57249.56224783 & \multicolumn{1}{c}{-} \\
        & 3230-7530 & B\&C/Asiago & 150-400 & -   & \multicolumn{1}{c}{-} & \multicolumn{1}{c}{-} \\
        & 3950-5780 & DADOS/OST   & 3500    & 2016-04-20 & 57498.94453704 & \multicolumn{1}{c}{-} \\
        & 5290-7140 & DADOS/OST   & 3500    & 2016-03-04 & 57451.02402778 & \multicolumn{1}{c}{-} \\
        & 14800-17800 & NICS/TNG  & 1150    & 2014-09-01 & 56901          & \multicolumn{1}{c}{-} \\
        & 19500-23400 & NICS/TNG  & 1250    & 2014-09-01 & 56901          & \multicolumn{1}{c}{-} \\
        HD\,100199  &  905-1187 & FUSE        & 20000   & 2000-03-24 & 51627.24236111 & \multicolumn{1}{c}{-} \\
        & 1150-1700 & STIS/\textsl{HST}    & 45800   & 2015-01-16 & 57038.99923304 & \multicolumn{1}{c}{-} \\
        & 3630-7170 &FEROS/ESO-2.2m&48000   & 2007-06-27 & 54278.00155806 & \multicolumn{1}{c}{-} \\
        & 3630-7170 &FEROS/ESO-2.2m&48000   & 2007-06-29 & 54280.96512831 & \multicolumn{1}{c}{-} \\
        \hline 
    \end{tabular}
\end{table*}

\begin{table*}[htbp]
    \caption{Photometry} 
    \label{table:photometry}
    \centering  
    \begin{tabular}{lSSSSSS}
        \hline\hline 
        & 
        \multicolumn{1}{c}{HD\,153919}  &
        \multicolumn{1}{c}{BD+60\,73}   &
        \multicolumn{1}{c}{LM\,Vel} &
        \multicolumn{1}{c}{HD\,306414}  &
        \multicolumn{1}{c}{BD+53\,2790} &
        \multicolumn{1}{c}{HD\,100199}   \rule[0mm]{0mm}{3.5mm} \\
        \hline  
        $U$\,(mag)   & 6.06\tablefootmark{a}  & 9.79\tablefootmark{b}  & 7.053\tablefootmark{c} & 10.12\tablefootmark{c}  &  9.42\tablefootmark{c}  & 7.351\tablefootmark{c} \rule[0mm]{0mm}{3.5mm} \\
        $B$\,(mag)   & 6.724\tablefootmark{d} & 10.21\tablefootmark{b} & 7.722\tablefootmark{d} & 10.52\tablefootmark{d}  & 10.11\tablefootmark{e}  & 8.19\tablefootmark{c}  \\
        $V$\,(mag)   & 6.543\tablefootmark{d} & 9.64\tablefootmark{b}  & 7.558\tablefootmark{d} & 10.11\tablefootmark{d}  &  9.84\tablefootmark{e}  & 8.187\tablefootmark{c}  \\
        $R$\,(mag)   & 6.43\tablefootmark{d}  & 9.31\tablefootmark{d}  & 7.47\tablefootmark{d}  &  9.84\tablefootmark{d}  &  9.64\tablefootmark{e}  & 8.18\tablefootmark{f}  \\
        $G$\,(mag)\tablefootmark{g}   & 6.38  & 9.4                    & 7.449                  &  9.703                  &  9.726                  & 8.176 \\
        $I$\,(mag)   & 5.93\tablefootmark{a}  & 9.072\tablefootmark{b} & \multicolumn{1}{c}{-}  &  9.41\tablefootmark{h}  &  9.43\tablefootmark{e}  & 8.22\tablefootmark{f}  \\
        $J$\,(mag)\tablefootmark{i}   & 5.744 & 8.389                  & 6.935                  &  8.548                  &  9.218                  & 8.048 \\
        $H$\,(mag)\tablefootmark{i}   & 5.639 & 8.265                  & 6.887                  &  8.340                  &  9.116                  & 8.067 \\
        $K_S$\,(mag)\tablefootmark{i} & 5.496 & 8.166                  & 6.808                  &  8.185                  &  9.038                  & 8.009 \\
        $W1$\,(mag)\tablefootmark{j}  & 5.36  & 8.104                  & 6.756                  &  8.043                  &  8.7                    & 8.063 \\
        $W2$\,(mag)\tablefootmark{j}  & 5.109 & 8.085                  & 6.687                  &  7.982                  &  8.562                  & 8.012 \\
        $W3$\,(mag)\tablefootmark{j}  & 4.927 & 7.994                  & 6.585                  &  7.807                  &  8.191                  & 7.625 \\
        $W4$\,(mag)\tablefootmark{j}  & 4.273 & 7.521                  & 6.207                  &  7.412                  &  7.9                    & 7.041 \\
        MSX6C A\,(Jy)& 0.6344\tablefootmark{k}& \multicolumn{1}{c}{-}  & \multicolumn{1}{c}{-}  & \multicolumn{1}{c}{-}   & \multicolumn{1}{c}{-}   & \multicolumn{1}{c}{-} \\
        \hline  
    \end{tabular}
    \tablefoot{
        \tablefoottext{a}{\citet{Morel1978}} 
        \tablefoottext{b}{\citet{Anderson2012}} 
        \tablefoottext{c}{\citet{Mermilliod2006}}    
        \tablefoottext{d}{\citet{Zacharias2004b}}
        \tablefoottext{e}{\citet{Reig2015}}
        \tablefoottext{f}{\citet{Monet2003}}
        \tablefoottext{g}{\citet{Gaia2016}}
        \tablefoottext{h}{\citet{DENIS2005}}
        \tablefoottext{i}{\citet{Cutri2003}}
        \tablefoottext{j}{\citet{Cutri2012a}}
        \tablefoottext{k}{\citet{Egan2003}}
    }
\end{table*}

\begin{table*}[htbp]
    \caption{X-ray measurements at times close to the UV observations} 
    \label{table:xray_data}
    \centering  
    \begin{tabular}{lcccSSS}
        \hline\hline 
        Identifier &
        ObsIDs &
        Observation mode &
        Observation date &
        \multicolumn{1}{c}{MJD} &
        \multicolumn{1}{c}{Phase} \rule[0mm]{0mm}{3.5mm} \\
         &
         &
         &
         &
        \multicolumn{1}{c}{(d)} &
         \\
        \hline  
        HD\,153919  & 00033631008 & WT & 2015-02-22 & 57075.17872458 & 0.141 \rule[0mm]{0mm}{4.0mm}\\        
        BD+60\,73   & 00032620025 & PC & 2015-01-01 & 57023.68082204 & 0.853 \\
        LM\,Vel     & 00037881103 & PC & 2015-07-08 & 57211.14901227 & 0.008 \\
                    & 00037881107 & PC & 2015-07-16 & 57219.12357639 & 0.901 \\
        HD\,306414  & 00030881043 & PC & 2015-08-16 & 57250.84636196 & \multicolumn{1}{c}{-} \\
        BD+53\,2790 & 00033914003 & WT & 2015-08-15 & 57249.21537077 & \multicolumn{1}{c}{-}	 \\
        HD\,100199  & 00035224007 & PC & 2015-01-15 & 57037.81825268 & \multicolumn{1}{c}{-} \\    
        \hline 
    \end{tabular}
\end{table*}

\begin{table*}[htbp]
    \caption{Atomic model used in the stellar atmosphere calculations} 
    \label{table:model_atoms}
    \centering  
    \begin{tabular}{lSS|lSS}
        \hline\hline 
        Ion &
        \multicolumn{1}{c}{Number of levels} &
        \multicolumn{1}{c}{Number of transitions}  &
        Ion & 
        \multicolumn{1}{c}{Number of levels} &
        \multicolumn{1}{c}{Number of transitions}  \rule[0mm]{0mm}{3.5mm} \\
        \hline  
        \ion{H}{i}    & 22 & 231 & \ion{Mg}{iii}                  & 43 & 903 \rule[0mm]{0mm}{4.0mm}  \\
        \ion{H}{ii}   & 1  &   0 & \ion{Mg}{iv}                   & 17 & 136 \\
        \ion{He}{i}   & 35 & 595 & \ion{Mg}{v}                    &  0 &   0 \\
        \ion{He}{ii}  & 26 & 325 & \ion{Mg}{vii}                  &  0 &   0 \\
        \ion{He}{iii} &  1 &   0 & \ion{Si}{ii}                   &  1 &   0 \\
        \ion{N}{ii}   & 38 & 703 & \ion{Si}{iii}                  & 24 & 276 \\
        \ion{N}{iii}  & 36 & 630 & \ion{Si}{iv}                   & 23 & 253 \\
        \ion{N}{iv}   & 38 & 703 & \ion{Si}{v}                    &  1 &   0 \\
        \ion{N}{v}    & 20 & 190 & \ion{P}{iv}                    & 12 &  66 \\
        \ion{N}{vi}   & 14 &  91 & \ion{P}{v}                     & 11 &  55 \\
        \ion{C}{ii}   & 32 & 496 & \ion{P}{vi}                    &  1 &   0 \\
        \ion{C}{iii}  & 40 & 780 & \ion{G}{ii}\tablefootmark{a}   &  1 &   0 \\
        \ion{C}{iv}   & 25 & 300 & \ion{G}{iii}\tablefootmark{a}  & 13 &  40 \\
        \ion{C}{v}    & 29 & 406 & \ion{G}{iv}\tablefootmark{a}   & 18 &  77 \\
        \ion{C}{vi}   & 15 & 105 & \ion{G}{v}\tablefootmark{a}    & 22 & 107 \\
        \ion{O}{ii}   & 37 & 666 & \ion{G}{vi}\tablefootmark{a}   & 29 & 194 \\
        \ion{O}{iii}  & 33 & 528 & \ion{G}{vii}\tablefootmark{a}  & 19 &  87 \\
        \ion{O}{iv}   & 29 & 406 & \ion{G}{viii}\tablefootmark{a} & 14 &  49 \\
        \ion{O}{v}    & 36 & 630 & \ion{G}{ix}\tablefootmark{a}   & 15 &  56 \\
        \ion{O}{vi}   & 16 & 120 & \ion{G}{x}\tablefootmark{a}    &  1 &  0  \\
        \ion{O}{vii}  &  0 &   0 & \ion{G}{xi}\tablefootmark{a}   &  0 &  0  \\
        \ion{O}{viii} &  0 &   0 & \ion{G}{xii}\tablefootmark{a}  &  0 &  0  \\
        \ion{S}{iii}  & 23 & 253 & \ion{G}{xiii}\tablefootmark{a} &  0 &  0  \\
        \ion{S}{iv}   & 11 &  55 & \ion{G}{xiv}\tablefootmark{a}  &  0 &  0  \\
        \ion{S}{v}    & 10 &  45 & \ion{G}{xv}\tablefootmark{a}   &  0 &  0  \\
        \ion{S}{vi}   &  1 &   0 & \ion{G}{xvi}\tablefootmark{a}  &  0 &  0  \\
        \ion{Mg}{i}   &  1 &   0 & \ion{G}{xvii}\tablefootmark{a} &  0 &  0  \\
        \ion{Mg}{ii}  & 32 & 496 &                                &    &     \\
        \hline 
    \end{tabular}
    \tablefoot{
        \tablefoottext{a}{G denotes a generic atom which incorporates the following 
            iron group elements \element{Fe}, \element{Sc}, \element{Ti}, \element{Cr}, 
            \element{Mn}, \element{Co}, and \element{Ni}. The  corresponding ions are 
            treated by means of a superlevel approach \citep[for details 
            see][]{Graefener2002}.}
    }
\end{table*}

\begin{table*}[htbp]
    \caption{Empirical stellar parameters, compared to the best-fitting single-star evolution model as interpolated with the BONNSAI tool} 
    \label{table:bonnsai}
    \centering  
\begin{tabular}{lcccccc}
        \hline\hline 
        &
        \multicolumn{2}{c}{HD\,153919} &
        \multicolumn{2}{c}{LM\,Vel} &
        \multicolumn{2}{c}{BD+53\,2790\tablefootmark{a}} \rule[0mm]{0mm}{3.5mm} \\
        & This study & BONNSAI & This study & BONNSAI & This study & BONNSAI \\
        \hline  
        $T_{\ast}$ (kK)  & 
        $35^{+2}_{-3}$ & $35^{+3}_{-3}$ & 
        $30^{+3}_{-3}$ & $29^{+3}_{-3}$ &
        $30^{+3}_{-3}$ & $31^{+3}_{-3}$         \rule[0mm]{0mm}{4.0mm}\\        
        $\log L$ ($L_\odot$)    &
        $5.7^{+0.1}_{-0.1}$ & $5.68^{+0.09}_{-0.08}$ &
        $5.3^{+0.1}_{-0.1}$ & $5.34^{+0.09}_{-0.09}$ &
        $4.9^{+0.1}_{-0.1}$ & $4.9^{+0.1}_{-0.1}$
        \rule[0mm]{0mm}{4.0mm}  \\
        $\log g_\ast$ (cm\,s$^{-2}$) & 
        $3.4^{+0.4}_{-0.4}$ & $3.5^{+0.2}_{-0.2}$ &
        $3.2^{+0.2}_{-0.2}$ & $3.3^{+0.2}_{-0.3}$  &
        $3.8^{+0.3}_{-0.5}$ & $3.8^{+0.2}_{-0.3}$ 
        \rule[0mm]{0mm}{4.0mm} \\
        $v \sin i$ (km\,s$^{-1}$)  & 
        $110^{+30}_{-50}$ & $100^{+40}_{-40}$ &
        $150^{+20}_{-20}$ & $150^{+20}_{-20}$ &
        $200^{+50}_{-50}$ & $200^{+45}_{-57}$ 
        \rule[0mm]{0mm}{4.0mm} \\
        $X_{\rm H}$  (mass fr.) &  
        $0.65^{+0.1}_{-0.2}$ & $0.72^{+0.00}_{-0.01}$ &
        $0.5^{+0.1}_{-0.1}$  & $0.72^{+0.00}_{-0.2}$  &
        \rule[0mm]{0mm}{4.0mm} \\
        $\log \dot{M}$ ($M_\odot \mathrm{yr}^{-1}$)  &  
        $-5.6^{+0.2}_{-0.3}$ & $-5.7^{+0.2}_{-0.2}$ &
        $-6.1^{+0.2}_{-0.2}$ & $-6.2^{+0.2}_{-0.2}$ &
        \rule[0mm]{0mm}{4.0mm} \\
\end{tabular}
         \tablefoot{
             \tablefoottext{a}{$X_{\rm H}$ and $\log \dot{M}$ not used as input for the BONNSAI tool.}
             \tablefoottext{b}{Parameter not used as input for the BONNSAI tool.}
         }
\end{table*}

\longtab[6]{
    \begin{longtable}{cccccccSS}
    \caption{Atmospheric structure of the model used to fit HD\,153919. The columns depict the index of the radius grid, the radius ($r$), the electron temperature ($T_\mathrm{e}$), the particle density ($N$), the electron density ($N_\mathrm{e}$), the Rosseland optical depth ($\tau_\mathrm\mathrm{Ross}$), the velocity ($v$), the velocity gradient ($\partial v / \partial r$), and the clumping factor ($D$).}
    \label{table:struct_hd153919}\\
        \hline\hline 
        \multicolumn{1}{c}{Depth}  & \multicolumn{1}{c}{$r-1$}      & \multicolumn{1}{c}{$T_\mathrm{e}$} & \multicolumn{1}{c}{$\log N$}                           & \multicolumn{1}{c}{$\log N_\mathrm{e}$}                    & \multicolumn{1}{c}{$\tau_\mathrm{Ross}$} & \multicolumn{1}{c}{$v$}        & \multicolumn{1}{c}{$\partial v / \partial r$}   & \multicolumn{1}{c}{$D$}  \rule[0mm]{0mm}{3.5mm} \\
        \multicolumn{1}{c}{index}  & \multicolumn{1}{c}{[$R_\ast$]} & \multicolumn{1}{c}{[K]}            & \multicolumn{1}{c}{$[\mathrm{Atoms} / \mathrm{cm}^3]$} & \multicolumn{1}{c}{$[\mathrm{Electrons} / \mathrm{cm}^3]$} & 				              & \multicolumn{1}{c}{[km\,/\,s]} & \multicolumn{1}{c}{[$\frac{\mathrm{km}\,/\,\mathrm{s}}{R_\ast}$]} &     \\
        \hline 
        \endfirsthead
        \caption{continued.}\\
        \hline
        \hline        
        \multicolumn{1}{c}{Depth}  & \multicolumn{1}{c}{$r-1$}      & \multicolumn{1}{c}{$T_\mathrm{e}$} & \multicolumn{1}{c}{$\log N$}                           & \multicolumn{1}{c}{$\log N_\mathrm{e}$}                    & \multicolumn{1}{c}{$\tau_\mathrm{Ross}$} & \multicolumn{1}{c}{$v$}        & \multicolumn{1}{c}{$\partial v / \partial r$}   & \multicolumn{1}{c}{$D$}  \rule[0mm]{0mm}{3.5mm} \\
        \multicolumn{1}{c}{index}  & \multicolumn{1}{c}{[$R_\ast$]} &  \multicolumn{1}{c}{[K]}           & \multicolumn{1}{c}{$[\mathrm{Atoms} / \mathrm{cm}^3]$} & \multicolumn{1}{c}{$[\mathrm{Electrons} / \mathrm{cm}^3]$} & 				              & \multicolumn{1}{c}{[km\,/\,s]} & \multicolumn{1}{c}{[$\frac{\mathrm{km}\,/\,\mathrm{s}}{R_\ast}$]} &   \\
        \hline
        \endhead
        \hline
        \endfoot
        \hline
        \endlastfoot 
        \rule[0mm]{0mm}{3.5mm} 
          1  &   99.0     & 16849. &  6.200 & 6.202 &  0.000000 &  1900.     &   0.51 & 20.00 \\
          2  &   78.6     & 16568. &  6.400 & 6.402 &  0.000044 &  1891.     &   0.51 & 20.00 \\
          3  &   65.0     & 16181. &  6.564 & 6.566 &  0.000087 &  1883.     &   0.76 & 20.00 \\
          4  &   54.5     & 15711. &  6.718 & 6.720 &  0.000136 &  1873.     &   1.10 & 20.00 \\
          5  &   45.1     & 15333. &  6.882 & 6.883 &  0.000199 &  1861.     &   1.61 & 20.00 \\
          6  &   36.6     & 15117. &  7.064 & 7.065 &  0.000284 &  1844.     &   2.43 & 20.00 \\
          7  &   28.7     & 15107. &  7.271 & 7.273 &  0.000407 &  1821.     &   3.70 & 20.00 \\
          8  &   22.9     & 15175. &  7.469 & 7.471 &  0.000555 &  1794.     &   5.48 & 20.00 \\
\end{longtable}
}

\longtab[7]{
    \begin{longtable}{cccccccSS}
        \caption{Same as Table\,\ref{table:struct_hd153919}, but for BD+60\,73}
        \label{table:struct_bd+6073}\\
        \hline\hline 
        \multicolumn{1}{c}{Depth}  & \multicolumn{1}{c}{$r-1$}      & \multicolumn{1}{c}{$T_\mathrm{e}$} & \multicolumn{1}{c}{$\log N$}                           & \multicolumn{1}{c}{$\log N_\mathrm{e}$}                    & \multicolumn{1}{c}{$\tau_\mathrm{Ross}$} & \multicolumn{1}{c}{$v$}        & \multicolumn{1}{c}{$\partial v / \partial r$}   & \multicolumn{1}{c}{$D$}  \rule[0mm]{0mm}{3.5mm} \\
        \multicolumn{1}{c}{index}  & \multicolumn{1}{c}{[$R_\ast$]} & \multicolumn{1}{c}{[K]}            & \multicolumn{1}{c}{$[\mathrm{Atoms} / \mathrm{cm}^3]$} & \multicolumn{1}{c}{$[\mathrm{Electrons} / \mathrm{cm}^3]$} & 				              & \multicolumn{1}{c}{[km\,/\,s]} & \multicolumn{1}{c}{[$\frac{\mathrm{km}\,/\,\mathrm{s}}{R_\ast}$]} &     \\
        \hline 
        \endfirsthead
        \caption{continued.}\\
        \hline
        \hline        
        \multicolumn{1}{c}{Depth}  & \multicolumn{1}{c}{$r-1$}      & \multicolumn{1}{c}{$T_\mathrm{e}$} & \multicolumn{1}{c}{$\log N$}                           & \multicolumn{1}{c}{$\log N_\mathrm{e}$}                    & \multicolumn{1}{c}{$\tau_\mathrm{Ross}$} & \multicolumn{1}{c}{$v$}        & \multicolumn{1}{c}{$\partial v / \partial r$}   & \multicolumn{1}{c}{$D$}  \rule[0mm]{0mm}{3.5mm} \\
        \multicolumn{1}{c}{index}  & \multicolumn{1}{c}{[$R_\ast$]} &  \multicolumn{1}{c}{[K]}           & \multicolumn{1}{c}{$[\mathrm{Atoms} / \mathrm{cm}^3]$} & \multicolumn{1}{c}{$[\mathrm{Electrons} / \mathrm{cm}^3]$} & 				              & \multicolumn{1}{c}{[km\,/\,s]} & \multicolumn{1}{c}{[$\frac{\mathrm{km}\,/\,\mathrm{s}}{R_\ast}$]} &   \\
        \hline
        \endhead
        \hline
        \endfoot
        \hline
        \endlastfoot 
        \rule[0mm]{0mm}{3.5mm} 
         1   &  99.0     &  10780. &  4.589 &  4.673 &  0.0000000 &  1100.     &   0.21 & 20.00 \\
         2   &  77.0     &  11061. &  4.807 &  4.891 &  0.0000011 &  1096.     &   0.21 & 20.00 \\
         3   &  62.4     &  11307. &  4.988 &  5.072 &  0.0000022 &  1092.     &   0.33 & 20.00 \\
         4   &  50.3     &  11610. &  5.175 &  5.259 &  0.0000036 &  1087.     &   0.51 & 20.00 \\
         5   &  40.0     &  11869. &  5.372 &  5.456 &  0.0000054 &  1081.     &   0.77 & 20.00 \\
         6   &  32.7     &  12085. &  5.544 &  5.629 &  0.0000074 &  1074.     &   1.18 & 20.00 \\
         7   &  26.6     &  12370. &  5.720 &  5.805 &  0.0000099 &  1065.     &   1.76 & 20.00 \\
         8   &  21.5     &  12672. &  5.904 &  5.989 &  0.0000131 &  1054.     &   2.56 & 20.00 \\
\end{longtable}
}

\longtab[8]{
    \begin{longtable}{cccccccSS}
        \caption{Same as Table\,\ref{table:struct_hd153919}, LM\,Vel}
        \label{table:struct_lm-vel}\\
        \hline\hline 
        \multicolumn{1}{c}{Depth}  & \multicolumn{1}{c}{$r-1$}      & \multicolumn{1}{c}{$T_\mathrm{e}$} & \multicolumn{1}{c}{$\log N$}                           & \multicolumn{1}{c}{$\log N_\mathrm{e}$}                    & \multicolumn{1}{c}{$\tau_\mathrm{Ross}$} & \multicolumn{1}{c}{$v$}        & \multicolumn{1}{c}{$\partial v / \partial r$}   & \multicolumn{1}{c}{$D$}  \rule[0mm]{0mm}{3.5mm} \\
        \multicolumn{1}{c}{index}  & \multicolumn{1}{c}{[$R_\ast$]} & \multicolumn{1}{c}{[K]}            & \multicolumn{1}{c}{$[\mathrm{Atoms} / \mathrm{cm}^3]$} & \multicolumn{1}{c}{$[\mathrm{Electrons} / \mathrm{cm}^3]$} & 				              & \multicolumn{1}{c}{[km\,/\,s]} & \multicolumn{1}{c}{[$\frac{\mathrm{km}\,/\,\mathrm{s}}{R_\ast}$]} &     \\
        \hline 
        \endfirsthead
        \caption{continued.}\\
        \hline
        \hline        
        \multicolumn{1}{c}{Depth}  & \multicolumn{1}{c}{$r-1$}      & \multicolumn{1}{c}{$T_\mathrm{e}$} & \multicolumn{1}{c}{$\log N$}                           & \multicolumn{1}{c}{$\log N_\mathrm{e}$}                    & \multicolumn{1}{c}{$\tau_\mathrm{Ross}$} & \multicolumn{1}{c}{$v$}        & \multicolumn{1}{c}{$\partial v / \partial r$}   & \multicolumn{1}{c}{$D$}  \rule[0mm]{0mm}{3.5mm} \\
        \multicolumn{1}{c}{index}  & \multicolumn{1}{c}{[$R_\ast$]} &  \multicolumn{1}{c}{[K]}           & \multicolumn{1}{c}{$[\mathrm{Atoms} / \mathrm{cm}^3]$} & \multicolumn{1}{c}{$[\mathrm{Electrons} / \mathrm{cm}^3]$} & 				              & \multicolumn{1}{c}{[km\,/\,s]} & \multicolumn{1}{c}{[$\frac{\mathrm{km}\,/\,\mathrm{s}}{R_\ast}$]} &   \\
        \hline
        \endhead
        \hline
        \endfoot
        \hline
        \endlastfoot 
        \rule[0mm]{0mm}{3.5mm} 
          1  &   99.0     &   12244.  &   5.776 &  5.777  &   0.000000 &  1900.     &    0.36 & 20.00 \\
          2  &   81.3     &   12377.  &   5.947 &  5.947  &   0.000011 &  1895.     &    0.36 & 20.00 \\
          3  &   66.8     &   12551.  &   6.117 &  6.117  &   0.000024 &  1888.     &    0.52 & 20.00 \\
          4  &   56.5     &   12679.  &   6.261 &  6.261  &   0.000037 &  1882.     &    0.74 & 20.00 \\
          5  &   48.0     &   12817.  &   6.401 &  6.402  &   0.000053 &  1874.     &    1.06 & 20.00 \\
          6  &   39.5     &   12960.  &   6.570 &  6.570  &   0.000075 &  1864.     &    1.55 & 20.00 \\
          7  &   31.3     &   13147.  &   6.769 &  6.770  &   0.000108 &  1849.     &    2.41 & 20.00 \\
          8  &   24.1     &   13368.  &   6.993 &  6.994  &   0.000156 &  1827.     &    3.80 & 20.00 \\
\end{longtable}
}

\longtab[9]{
    \begin{longtable}{cccccccSS}
        \caption{Same as Table\,\ref{table:struct_hd153919}, HD\,306414}
        \label{table:struct_hd306414}\\
        \hline\hline 
        \multicolumn{1}{c}{Depth}  & \multicolumn{1}{c}{$r-1$}      & \multicolumn{1}{c}{$T_\mathrm{e}$} & \multicolumn{1}{c}{$\log N$}                           & \multicolumn{1}{c}{$\log N_\mathrm{e}$}                    & \multicolumn{1}{c}{$\tau_\mathrm{Ross}$} & \multicolumn{1}{c}{$v$}        & \multicolumn{1}{c}{$\partial v / \partial r$}   & \multicolumn{1}{c}{$D$}  \rule[0mm]{0mm}{3.5mm} \\
        \multicolumn{1}{c}{index}  & \multicolumn{1}{c}{[$R_\ast$]} & \multicolumn{1}{c}{[K]}            & \multicolumn{1}{c}{$[\mathrm{Atoms} / \mathrm{cm}^3]$} & \multicolumn{1}{c}{$[\mathrm{Electrons} / \mathrm{cm}^3]$} & 				              & \multicolumn{1}{c}{[km\,/\,s]} & \multicolumn{1}{c}{[$\frac{\mathrm{km}\,/\,\mathrm{s}}{R_\ast}$]} &     \\
        \hline 
        \endfirsthead
        \caption{continued.}\\
        \hline
        \hline        
        \multicolumn{1}{c}{Depth}  & \multicolumn{1}{c}{$r-1$}      & \multicolumn{1}{c}{$T_\mathrm{e}$} & \multicolumn{1}{c}{$\log N$}                           & \multicolumn{1}{c}{$\log N_\mathrm{e}$}                    & \multicolumn{1}{c}{$\tau_\mathrm{Ross}$} & \multicolumn{1}{c}{$v$}        & \multicolumn{1}{c}{$\partial v / \partial r$}   & \multicolumn{1}{c}{$D$}  \rule[0mm]{0mm}{3.5mm} \\
        \multicolumn{1}{c}{index}  & \multicolumn{1}{c}{[$R_\ast$]} &  \multicolumn{1}{c}{[K]}           & \multicolumn{1}{c}{$[\mathrm{Atoms} / \mathrm{cm}^3]$} & \multicolumn{1}{c}{$[\mathrm{Electrons} / \mathrm{cm}^3]$} & 				              & \multicolumn{1}{c}{[km\,/\,s]} & \multicolumn{1}{c}{[$\frac{\mathrm{km}\,/\,\mathrm{s}}{R_\ast}$]} &   \\
        \hline
        \endhead
        \hline
        \endfoot
        \hline
        \endlastfoot 
        \rule[0mm]{0mm}{3.5mm} 
         1 &  99.0     & 11408. &  5.354 &  5.358 &  0.0000000 &  800.0     &   0.33 & 20.00 \\
         2 &  77.9     & 11693. &  5.563 &  5.568 &  0.0000086 &  794.5     &   0.33 & 20.00 \\
         3 &  61.0     & 12069. &  5.777 &  5.782 &  0.0000120 &  787.3     &   0.52 & 20.00 \\
         4 &  49.0     & 12371. &  5.968 &  5.973 &  0.0000328 &  779.4     &   0.79 & 20.00 \\
         5 &  40.5     & 12609. &  6.135 &  6.140 &  0.0000465 &  771.1     &   1.15 & 20.00 \\
         6 &  34.4     & 12800. &  6.276 &  6.282 &  0.0000603 &  762.7     &   1.59 & 20.00 \\
         7 &  29.5     & 12977. &  6.413 &  6.419 &  0.0000760 &  753.5     &   2.21 & 20.00 \\
         8 &  24.5     & 13177. &  6.576 &  6.582 &  0.0000984 &  740.7     &   3.15 & 20.00 \\
\end{longtable}
}

\longtab[10]{
    \begin{longtable}{cccccccSS}
        \caption{Same as Table\,\ref{table:struct_hd153919}, BD+53\,2790}
        \label{table:struct_bd+532790}\\
        \hline\hline 
        \multicolumn{1}{c}{Depth}  & \multicolumn{1}{c}{$r-1$}      & \multicolumn{1}{c}{$T_\mathrm{e}$} & \multicolumn{1}{c}{$\log N$}                           & \multicolumn{1}{c}{$\log N_\mathrm{e}$}                    & \multicolumn{1}{c}{$\tau_\mathrm{Ross}$} & \multicolumn{1}{c}{$v$}        & \multicolumn{1}{c}{$\partial v / \partial r$}   & \multicolumn{1}{c}{$D$}  \rule[0mm]{0mm}{3.5mm} \\
        \multicolumn{1}{c}{index}  & \multicolumn{1}{c}{[$R_\ast$]} & \multicolumn{1}{c}{[K]}            & \multicolumn{1}{c}{$[\mathrm{Atoms} / \mathrm{cm}^3]$} & \multicolumn{1}{c}{$[\mathrm{Electrons} / \mathrm{cm}^3]$} & 				              & \multicolumn{1}{c}{[km\,/\,s]} & \multicolumn{1}{c}{[$\frac{\mathrm{km}\,/\,\mathrm{s}}{R_\ast}$]} &     \\
        \hline 
        \endfirsthead
        \caption{continued.}\\
        \hline
        \hline        
        \multicolumn{1}{c}{Depth}  & \multicolumn{1}{c}{$r-1$}      & \multicolumn{1}{c}{$T_\mathrm{e}$} & \multicolumn{1}{c}{$\log N$}                           & \multicolumn{1}{c}{$\log N_\mathrm{e}$}                    & \multicolumn{1}{c}{$\tau_\mathrm{Ross}$} & \multicolumn{1}{c}{$v$}        & \multicolumn{1}{c}{$\partial v / \partial r$}   & \multicolumn{1}{c}{$D$}  \rule[0mm]{0mm}{3.5mm} \\
        \multicolumn{1}{c}{index}  & \multicolumn{1}{c}{[$R_\ast$]} &  \multicolumn{1}{c}{[K]}           & \multicolumn{1}{c}{$[\mathrm{Atoms} / \mathrm{cm}^3]$} & \multicolumn{1}{c}{$[\mathrm{Electrons} / \mathrm{cm}^3]$} & 				              & \multicolumn{1}{c}{[km\,/\,s]} & \multicolumn{1}{c}{[$\frac{\mathrm{km}\,/\,\mathrm{s}}{R_\ast}$]} &   \\
        \hline
        \endhead
        \hline
        \endfoot
        \hline
        \endlastfoot 
        \rule[0mm]{0mm}{3.5mm} 
         1 &  99.0     & 11737. &   5.542 &  5.569 &  0.0000000 &  400.0     &    0.06 & 8.58 \\
         2 &  78.1     & 12020. &   5.746 &  5.775 &  0.0000050 &  398.9     &    0.06 & 8.57 \\
         3 &  61.3     & 12482. &   5.955 &  5.984 &  0.0000116 &  397.6     &    0.10 & 8.55 \\
         4 &  49.4     & 12842. &   6.141 &  6.170 &  0.0000189 &  396.0     &    0.16 & 8.54 \\
         5 &  39.5     & 13227. &   6.333 &  6.363 &  0.0000283 &  394.1     &    0.25 & 8.51 \\
         6 &  31.1     & 13612. &   6.538 &  6.568 &  0.0000410 &  391.4     &    0.38 & 8.48 \\
         7 &  25.2     & 13934. &   6.720 &  6.749 &  0.0000549 &  388.6     &    0.60 & 8.44 \\
         8 &  20.1     & 14242. &   6.912 &  6.942 &  0.0000733 &  384.8     &    0.96 & 8.39 \\
\end{longtable}
}

\longtab[11]{
    \begin{longtable}{cccccccSS}
        \caption{Same as Table\,\ref{table:struct_hd153919}, HD\,100199.}
        \label{table:struct_hd100199}\\
        \hline\hline 
        \multicolumn{1}{c}{Depth}  & \multicolumn{1}{c}{$r-1$}      & \multicolumn{1}{c}{$T_\mathrm{e}$} & \multicolumn{1}{c}{$\log N$}                           & \multicolumn{1}{c}{$\log N_\mathrm{e}$}                    & \multicolumn{1}{c}{$\tau_\mathrm{Ross}$} & \multicolumn{1}{c}{$v$}        & \multicolumn{1}{c}{$\partial v / \partial r$}   & \multicolumn{1}{c}{$D$}  \rule[0mm]{0mm}{3.5mm} \\
        \multicolumn{1}{c}{index}  & \multicolumn{1}{c}{[$R_\ast$]} & \multicolumn{1}{c}{[K]}            & \multicolumn{1}{c}{$[\mathrm{Atoms} / \mathrm{cm}^3]$} & \multicolumn{1}{c}{$[\mathrm{Electrons} / \mathrm{cm}^3]$} & 				              & \multicolumn{1}{c}{[km\,/\,s]} & \multicolumn{1}{c}{[$\frac{\mathrm{km}\,/\,\mathrm{s}}{R_\ast}$]} &     \\
        \hline 
        \endfirsthead
        \caption{continued.}\\
        \hline
        \hline        
        \multicolumn{1}{c}{Depth}  & \multicolumn{1}{c}{$r-1$}      & \multicolumn{1}{c}{$T_\mathrm{e}$} & \multicolumn{1}{c}{$\log N$}                           & \multicolumn{1}{c}{$\log N_\mathrm{e}$}                    & \multicolumn{1}{c}{$\tau_\mathrm{Ross}$} & \multicolumn{1}{c}{$v$}        & \multicolumn{1}{c}{$\partial v / \partial r$}   & \multicolumn{1}{c}{$D$}  \rule[0mm]{0mm}{3.5mm} \\
        \multicolumn{1}{c}{index}  & \multicolumn{1}{c}{[$R_\ast$]} &  \multicolumn{1}{c}{[K]}           & \multicolumn{1}{c}{$[\mathrm{Atoms} / \mathrm{cm}^3]$} & \multicolumn{1}{c}{$[\mathrm{Electrons} / \mathrm{cm}^3]$} & 				              & \multicolumn{1}{c}{[km\,/\,s]} & \multicolumn{1}{c}{[$\frac{\mathrm{km}\,/\,\mathrm{s}}{R_\ast}$]} &   \\
        \hline
        \endhead
        \hline
        \endfoot
        \hline
        \endlastfoot 
        \rule[0mm]{0mm}{3.5mm} 
          1 &    99.0      & 12487. &  4.448 &  4.479  &  0.00000000 &  1500.     &    0.19 & 10.00 \\       
          2 &    76.4      & 12519. &  4.672 &  4.703  &  0.00000026 &  1496.     &    0.19 & 10.00 \\       
          3 &    61.5      & 12556. &  4.859 &  4.890  &  0.00000054 &  1493.     &    0.31 & 10.00 \\       
          4 &    49.0      & 12603. &  5.054 &  5.085  &  0.00000089 &  1488.     &    0.49 & 10.00 \\       
          5 &    38.4      & 12660. &  5.261 &  5.293  &  0.00000138 &  1481.     &    0.76 & 10.00 \\       
          6 &    31.0      & 12720. &  5.446 &  5.477  &  0.00000191 &  1474.     &    1.20 & 10.00 \\       
          7 &    24.8      & 12783. &  5.636 &  5.668  &  0.00000260 &  1465.     &    1.97 & 10.00 \\       
          8 &    18.5      & 12868. &  5.884 &  5.915  &  0.00000376 &  1449.     &    3.36 & 10.00 \\
\end{longtable}
}

\clearpage

\section{Comments on the individual stars}
\label{sec:comments}

\paragraph{HD\,153919} (4U\,1700-37): the donor star in this persistent HMXB has the earliest spectral-type in our sample, exhibiting prominent emission lines in its spectrum.
Based on our spectra and our spectral analysis we would classify this donor star as O6\,If/WN9, in contrast to the O6\,Iafpe classification assigned by \citet{Sota2014}. We would assign this different spectral type, since from our perspective this object is actually evolving from an Of to a WN star, in contrast to what is discussed by \citet{Sota2014} for the O6\,Iafpe classification. In this sense, the O6\,If/WN9 category would be an extension of the Of/WN class to cooler temperatures in reminiscence of the old ``cool slash'' category. From our perspective, an O6\,If/WN9 classification would be more suitable also in representation of the wind parameters of this object, which point to an object that is on its way to the WR stage. The derived mass-loss rate is compatible with that of other Of/WN stars \citep{Hainich2014}. 

The basic stellar parameters derived in this work are in good agreement with the previous results obtained by \citet{Clark2002}. The mass-loss rate derived by means of our models accounting for wind inhomogeneities is almost a factor of four lower than the value obtained by \citet{Clark2002} with unclumped models. The latter authors already have noted that moderate wind clumping would reduce their derived mass-loss rate. Taking into account the uncertainties of the individual studies, this brings the two works into agreement.
We note that the terminal velocity determined from our \textsl{HST} spectrum is slightly higher (by 150\,$\mathrm{km}\,\mathrm{s}^{-1}$) than obtained by \citet{Clark2002}. 

Interestingly, the hydrogen abundance deduced from our spectral fit coincides (within the uncertainties) with the value assumed by \citet{Clark2002}. While we also derived a supersolar nitrogen abundance, it is a factor of three lower compared to the value determined by \citet{Clark2002}. The carbon and oxygen abundances are in a better agreement. Like \citet{Clark2002}, we determine a solar carbon abundance
and a oxygen abundance of about $0.5\,X_{\mathrm{O}, \odot}$.

\paragraph{BD+60\,73} (IGR\,J00370+6122): according to \citet{Gonzalez-Galan2014}, this system is intermediate between a persistent HMXB and an ``intermediate'' SFXT because of its exceptional X-ray properties. 
In contrast to almost all other donor stars in our sample, a micro turbulence velocity of $\xi = 17^{+2}_{-2}\,\mathrm{km}\,\mathrm{s}^{-1}$ is required to achieve a satisfying fit.
Most of the stellar parameters we deduce for BD+60\,73 agree very well with the results by \citet{Gonzalez-Galan2014}. While these authors assume a wind-strength \mbox{$Q$-parameter} of \mbox{$\log Q = -13.0$}, our detailed wind analysis results in a value that almost a factor of three higher.
Also the derived abundances partly differ. The carbon and nitrogen abundances are a factor of about 1.5 higher in our study than the results presented by \citet{Gonzalez-Galan2014}, while our oxygen abundance is lower by the same factor. 
The deviation is the highest for the magnesium abundances, which is twice as high in our study compared to their value. The derived silicon abundances are approximately compatible. 
The same holds for the hydrogen abundance, which is only a few percent lower in this work.

In the fit shown in Fig.\,\ref{fig:bd+6073}, the model obviously falls short to reproduce the resonance doublets of \ion{N}{v}\,$\lambda\lambda$\,1239, 1243 and \ion{C}{iv}\,$\lambda\lambda$\,5801, 5812 with the observed strength. This model has been calculated with an X-ray irradiation that is consistent with the \textsl{Swift} observation. However, if we adopt an approximately 70 times higher X-ray irradiation, those resonance doublets perfectly match the observation, as demonstrated in Fig.\,\ref{fig:bd+6073_2}. Obviously, the stronger X-ray field causes sufficient photo- and Auger ionization to populate the N\,{\sc v} and C\,{\sc iv} ground states.

At this point, we have to realize that the X-ray measurement with \textsl{Swift} was not strictly simultaneous to our HST exposure, but was taken 4.5\,h later for technical reasons. Thus, given the X-ray variability of this target, we conclude that at the exact time of the HST observation the X-ray irradiation was somewhat enhanced due to some kind of flare.

\paragraph{LM\,Vel} (IGR\,J08408-4503): to our knowledge, the spectral analysis presented here is the first one for LM\,Vel. 
Although the overall spectral fit represents the observed spectrum very well, we are not able to achieve a satisfactory fit of the \ion{C}{iii} line at 1245\,\AA, which is stronger in the model compared to the observation. This might be a result of the neglection of macro clumping in our analysis, which in turn might imply an underestimation of the mass-loss rate \citep{Oskinova2007}.

Similar to BD+60\,73,
the model that has been calculated with an X-ray irradiation, which is consistent with the \textsl{Swift} data, falls short to reproduce the \ion{N}{v}\,$\lambda\lambda$\,1239, 1243 doublet (see Fig.\,\ref{fig:lm_vel}). Those models that are able to reproduce this doublet to a satisfactory level (see Fig.\,\ref{fig:lm_vel_2}) require an X-ray flux that is roughly 300 times higher than measured by the \textsl{Swift} observations. For technical reasons, the \textsl{Swift} data was taken 6.2\,h earlier than the HST data. Thus, this X-ray transient might experienced an X-ray outburst during our HST observations. 

As for HD\,153919, we find that this donor star is hydrogen and oxygen depleted, while the carbon abundance is solar and the nitrogen abundance is supersolar.

\paragraph{BD+53\,2790} (4U\,2206+54): unfortunately, we only have low resolution optical spectra with a low S/N at hand for this object, which is one reason for the relatively large error margins for some of the stellar parameters listed in Table\,\ref{table:parameters}. Nevertheless, these spectra clearly show a double peaked H$\alpha$ emission line, as typical for the decretion disks of Be- and Oe-type stars. The same spectral characteristic is posed by the hydrogen lines in the H- and K-band spectra shown in Fig.\,\ref{fig:bd+532790}. However, \citet[][see also \citealt{Negueruela2001}]{Blay2006} argue that this star does not fulfill all criteria of a classical Be-type star, but is rather a peculiar O9.5\,V star. While the donor star in this system is analyzed in this work, it is not considered in the discussion section of this paper because of its unclear HMXB type. 

Since our atmosphere models are restricted to spherical symmetry, we cannot account for asymmetries caused by the high rotational velocities of Be- and Oe-type stars, such as oblateness or decretion disks. Nevertheless, an adequate spectral fit can be achieved for most parts of the observed spectrum, with the exception of the hydrogen lines that are filled by the emission from the decretion disks. We also note that the width of the emission peaks of the resonance lines of \ion{C}{iv} and \ion{N}{v} in the UV cannot be reproduced completely by our model, most likely because of asymmetries in the wind of this star.
Since the \element{H}$\alpha$ line is dominated by emission from the decretion disk, this line cannot be used to constrain the clumping within the donor star atmosphere and wind. Therefore, we assume a clumping factor of $D = 10$.

\paragraph{HD\,306414} (IGR\,J11215-5952): this system is one of the SFXTs in our sample. Since it was not detected in our \textsl{Swift} observations, we had to assume a certain X-ray flux to proceed with the atmosphere model fits. 

Massive stars are inherent X-ray sources because of their winds that exhibit an intrinsic instability. This so-called line-driven wind instability \citep{Lucy1970} gives rise to wind inhomogeneities as well as shocks that can produce X-rays  \citep[e.g.,][]{Feldmeier1997,Runacres2002}. The intrinsic X-ray flux of massive stars is proportional to their stellar luminosity with $L_X\,/\,L \approx 10^{-7}$ \citep{Pallavicini1981}.

In the atmosphere model for this source, we therefore approximated the X-ray flux by two components. For the first one, we used a relatively soft X-ray continuum corresponding to an X-ray temperature of $T_X = 3 \times 10^{6}\,\mathrm{K}$. This component was inserted at a radius of $1.5\,R_\ast$, while the corresponding filling factor was adjusted such that $L_X \approx 10^{-7}\,L$ is produced. To model the contribution of the NS to the X-ray emission, a second X-ray continuum with an X-ray temperature of $T_X = 3 \times 10^{7}\,\mathrm{K}$ was injected at the position of the NS. The filling factor for this component was chosen such that the UV observations are reproduced best by the model, while ensuring that the total X-ray flux is below the detection limit of \textsl{Swift}.

The donor star has previously been analyzed by \citet{Lorenzo2014}. While we obtain a slightly higher stellar temperature and surface gravity as the latter authors, the luminosity derived in our analysis is 0.2\,dex lower even after accounting for the difference in the assumed distance. The reason for this discrepancy might be the different reddening estimates. While in our case the reddening is derived from an SED fit spanning from UV to infrared data, the estimate conducted by \citet{Lorenzo2014} is solely based on optical and IR photometry, leading to a significantly higher $R_V$ value of 4.2 and a slightly lower $E_{B-V} = 0.7$. Assuming these values for our model SED does not result in a satisfactory fit, providing confidence to our solution. 
The lower luminosity obtained from our analysis in comparison to that derived by \citet{Lorenzo2014} also entails a spectroscopic mass that is about 30\,\% lower. 

Our spectral analysis based on UV and optical data also results in a significantly lower mass-loss rate than determined by \citet{Lorenzo2014} solely on the basis of optical spectra. This discrepancy in the derived mass-loss rate can be in large part attributed to the neglect of wind inhomogeneities in the spectral analysis by \citet{Lorenzo2014}. If we scale the mass-loss rate determined by \citet{Lorenzo2014} according to the clumping factor ($D = 20$) derived in this work, the discrepancy nearly vanishes. 

The hydrogen, oxygen, and magnesium abundances determine by our analysis agree very well with the ones obtained by \citet{Lorenzo2014}. The carbon abundances coincide on a 20\,\% level, while the nitrogen and silicon abundance are higher by 30\,\% and 40\,\%, respectively, in our study compared to those derived by \citet{Lorenzo2014}.
These deviations might be a result of different micro turbulence velocities assumed in the spectral analyses. Unfortunately, \citet{Lorenzo2014} do not specify the micro turbulence velocity they assume. However, a value slightly different to the $\xi = 20^{+5}_{-5}\,\mathrm{km}\,\mathrm{s}^{-1}$ required by our analysis might explain the differences in the abundance measurements.

\paragraph{HD\,100199} (IGR\,J11305-6256): in this work we present the first spectral analysis of this Be X-ray binary. The same restrictions as outlined for BD+53\,2790 apply to the spectral modeling of HD\,100199. Overall, an excellent fit quality could be achieved with the exception of the line cores of the hydrogen lines in the optical. As for BD+53\,2790, we are not able to constrain the clumping and assume $D = 10$.

\clearpage

\section{Spectral fits}
\label{sect:spectra}

 \begin{figure*}
   \centering
   \includegraphics[angle=90,width=0.9\textwidth,page=1]{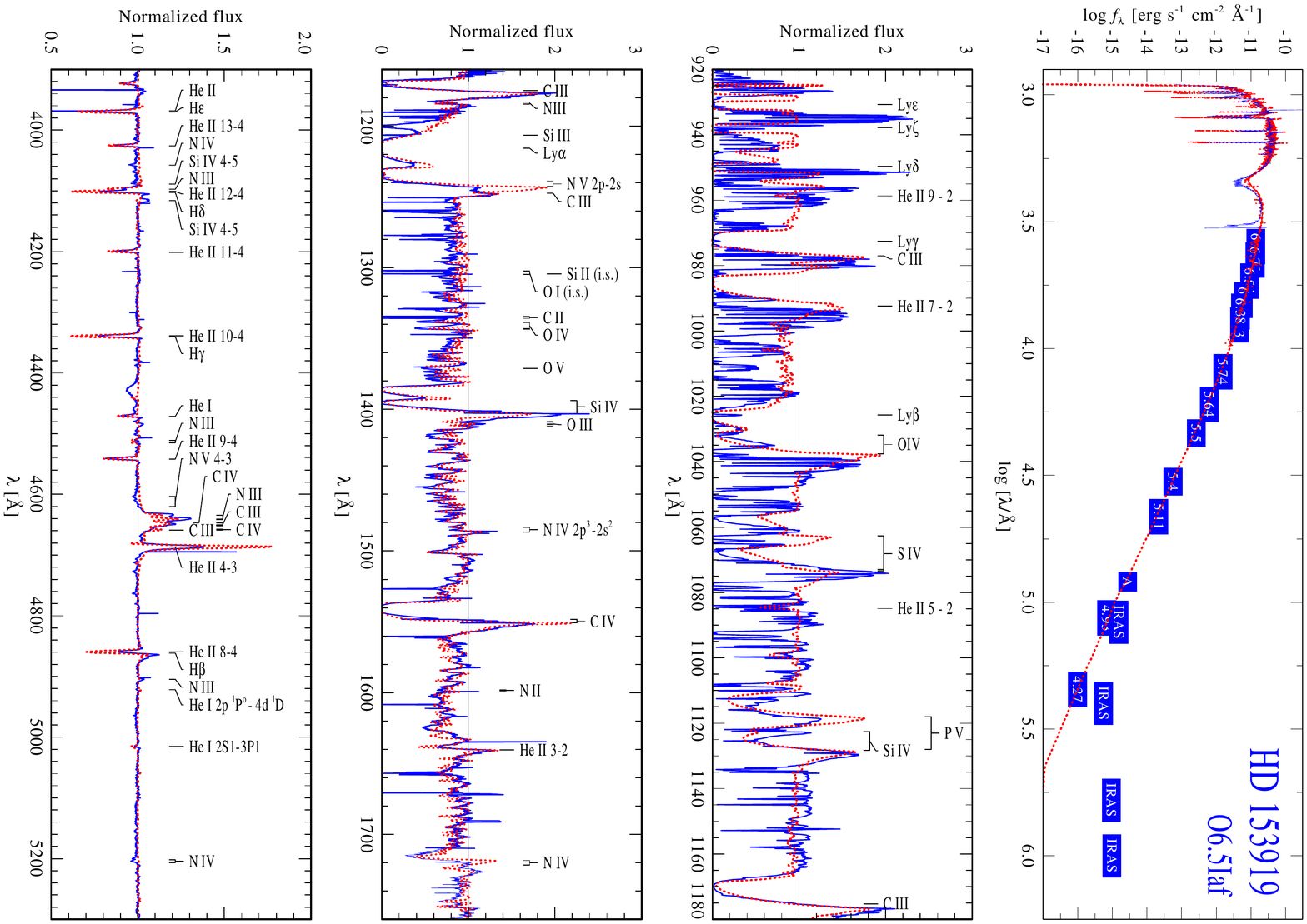}
   \caption{Spectral fit of HD\,153919. The observations are shown as blue continuous lines (spectra) and blue boxes (photometry). The best fitting model is overplotted by a dashed red line. Note that the observed far UV spectrum (FUSE) is heavily contaminated by interstellar abortion lines, mostly originating from \element{H}$_2$.
   }
   \label{fig:hd153919}
 \end{figure*}

\clearpage

\setcounter{figure}{0}

 \begin{figure*}
   \centering  
   \includegraphics[angle=90,width=0.92\textwidth,page=2,,trim=590 0 0 0,clip]{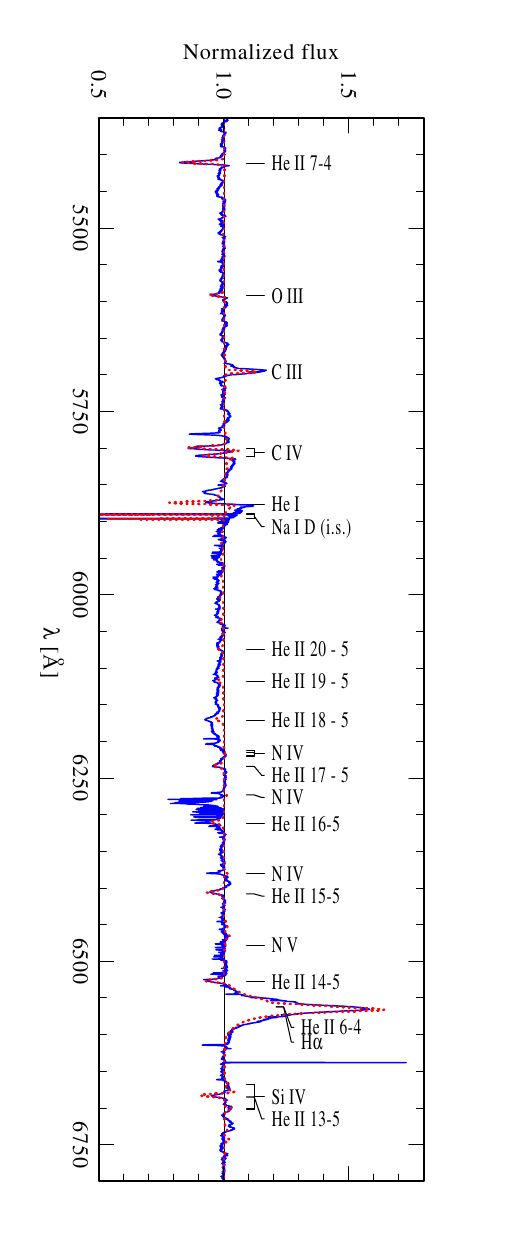}
   \caption{continued.}
 \end{figure*}

\begin{figure*}
    \centering
    \includegraphics[angle=90,width=0.92\textwidth,page=1]{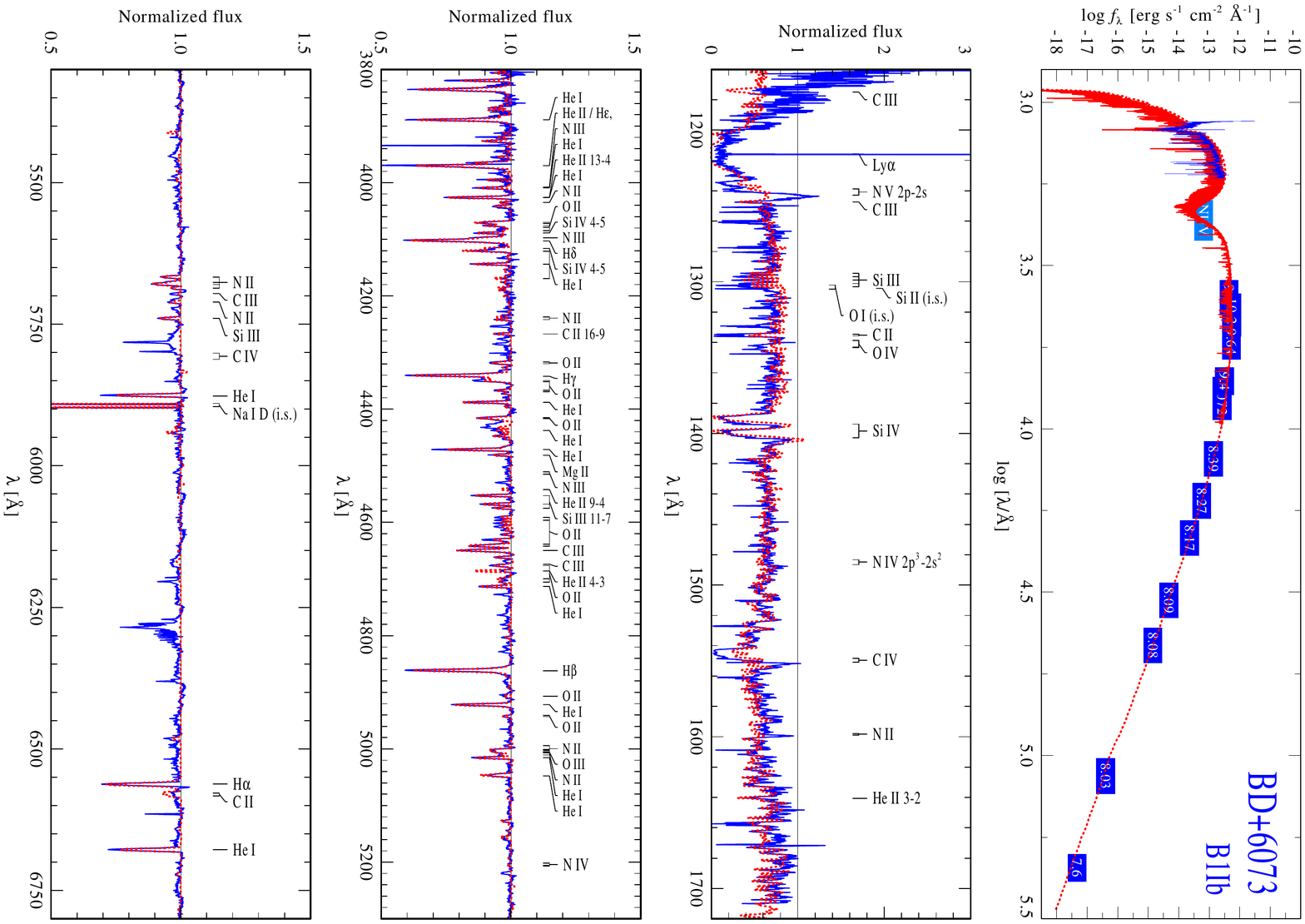}
    \caption{Same as Fig.\,\ref{fig:hd153919}, but for BD+60\,73.
    }
    \label{fig:bd+6073}
\end{figure*}

\begin{figure*}
\centering
\includegraphics[angle=90,width=0.92\textwidth]{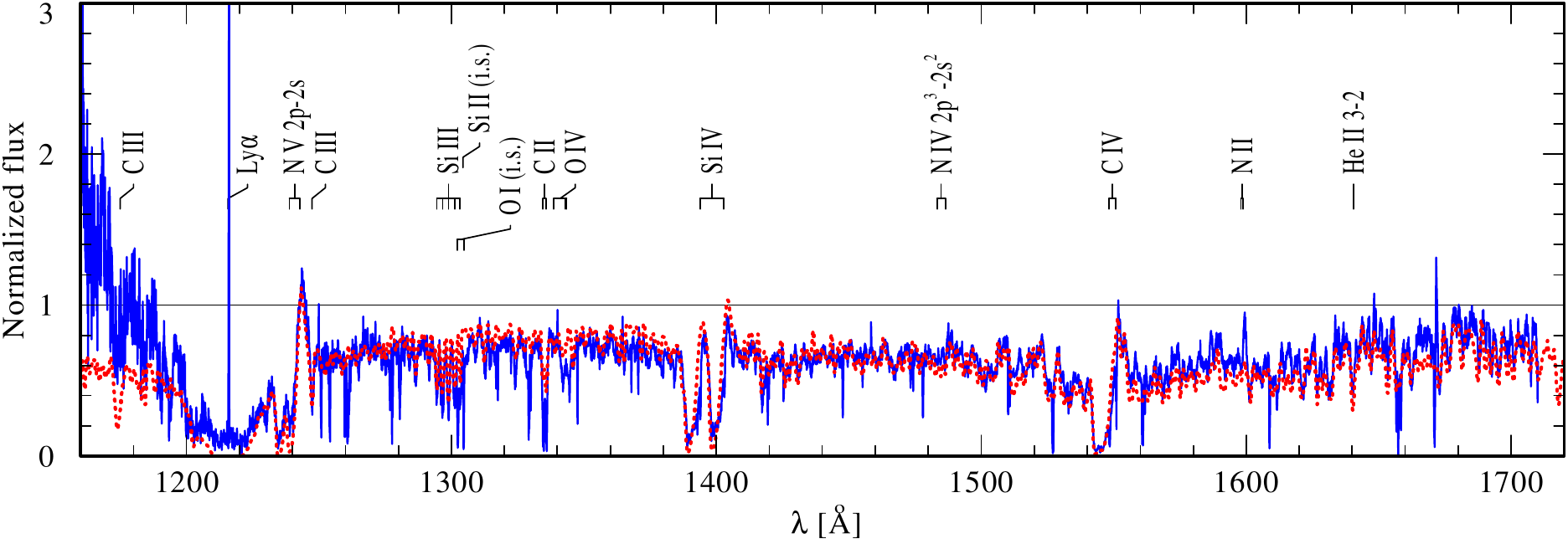}
\caption{BD+60\,73: Alternative UV-line fit (cf.\ Fig.\,\ref{fig:bd+6073}, second panel). For the model shown here, an approximately 70 times higher X-ray irradiation has been adopted, which brings the resonance doublets of \ion{N}{v}\,$\lambda\lambda$\,1239, 1243 and \ion{C}{iv}\,$\lambda\lambda$\,5801, 5812 to the observed strength (see text in Appendix\,\ref{sec:comments}).}
\label{fig:bd+6073_2}
\end{figure*}

\begin{figure*}
    \centering
    \includegraphics[angle=90,width=0.92\textwidth,page=1]{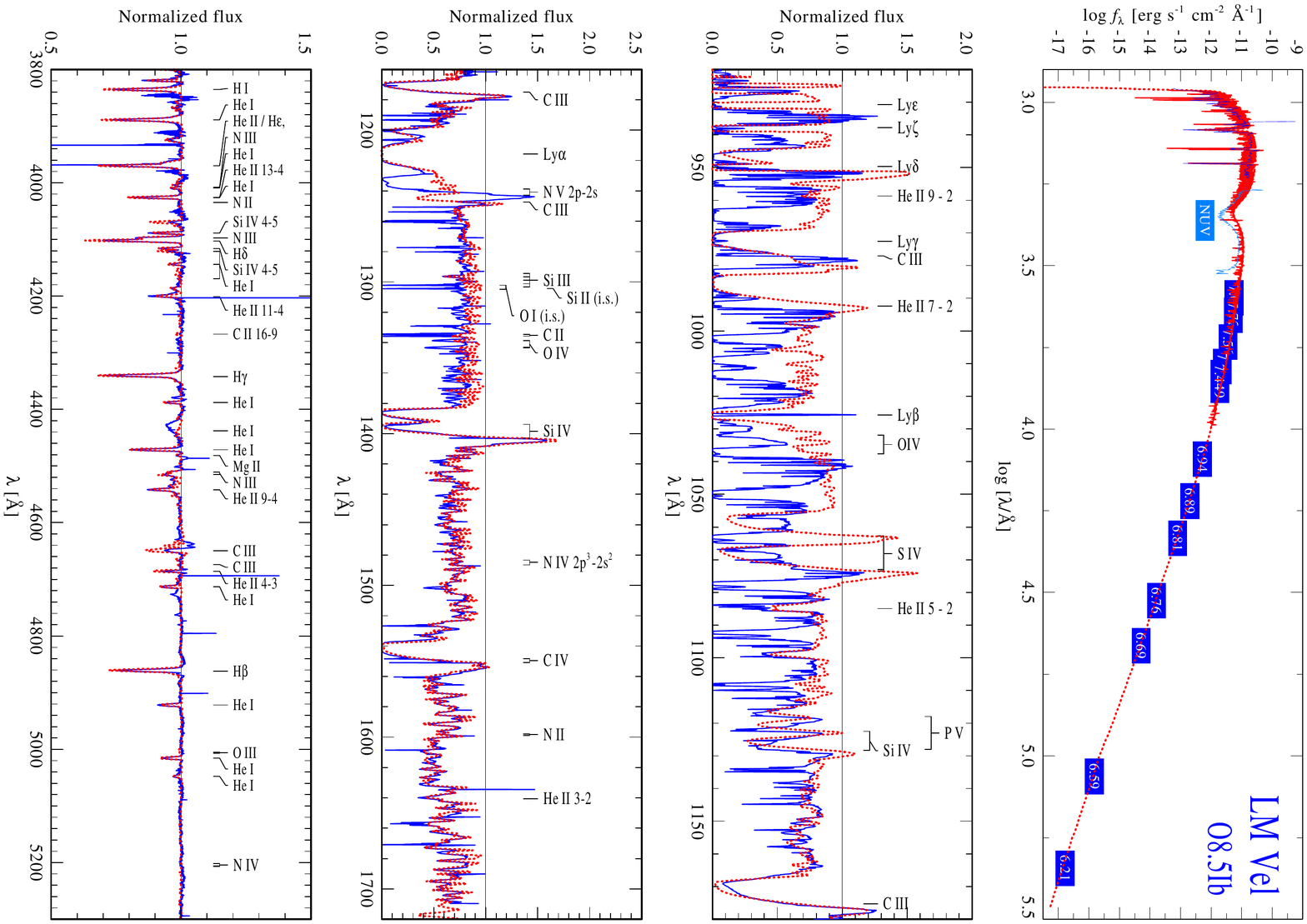}
    \caption{Same as Fig.\,\ref{fig:hd153919}, but for LM\,Vel.
    }
    \label{fig:lm_vel}
\end{figure*}

\clearpage

\addtocounter{figure}{-1}

\begin{figure*}
    \centering  
    \includegraphics[angle=90,width=0.92\textwidth,page=2,,trim=590 0 0 0,clip]{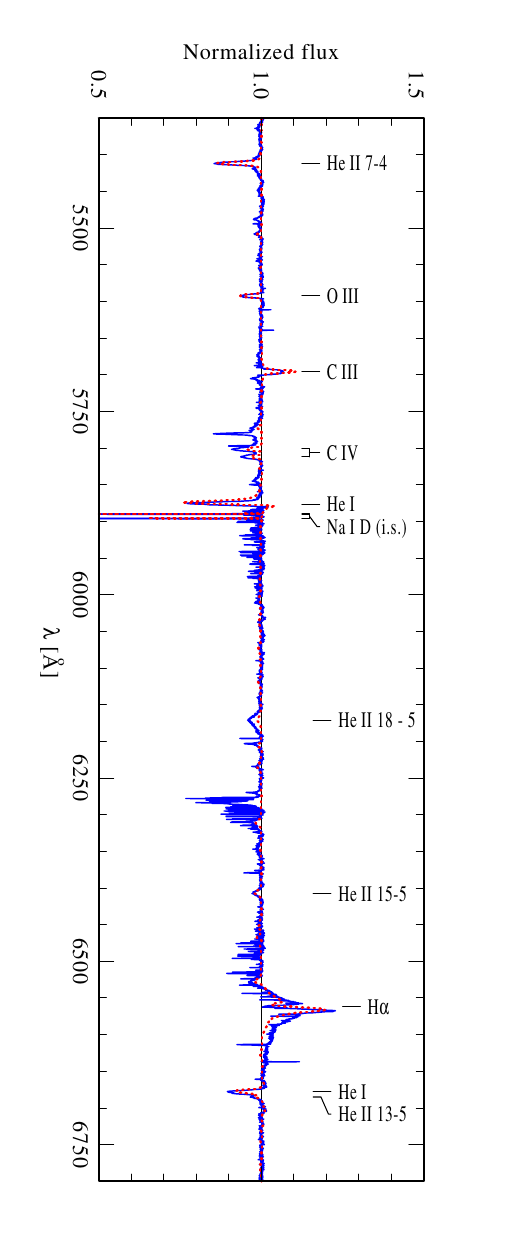}
    \caption{continued.}
\end{figure*}

\begin{figure*}
    \centering
    \includegraphics[angle=90,width=0.92\textwidth]{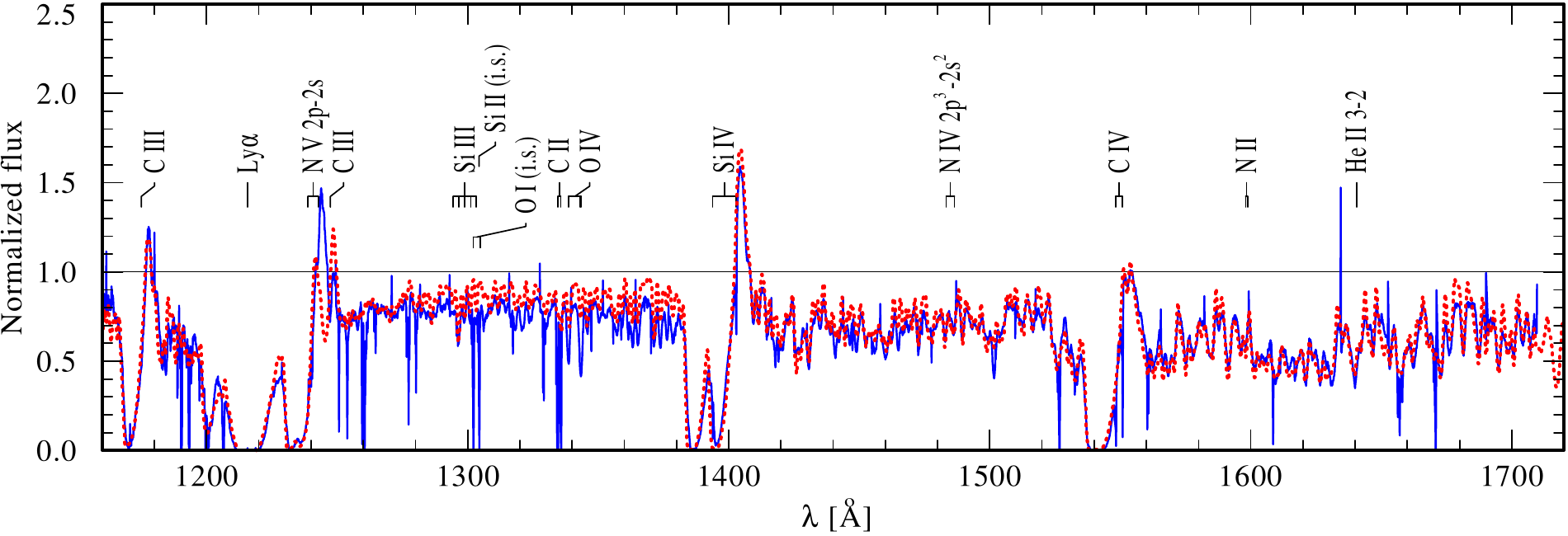}
    \caption{LM\,Vel: alternative UV-line fit (cf.\ Fig.\,\ref{fig:lm_vel}, second panel). For the model shown here, a roughly 300 times higher X-ray irradiation has been adopted, which brings the resonance doublet of \ion{N}{v}\,$\lambda\lambda$\,1239, 1243 almost to the observed strength (see text in Appendix\,\ref{sec:comments}). The remaining discrepancy is because of the \ion{C}{iii} line at 1245\,\AA\  (see Appendix\,\ref{sec:comments} for details).}
    \label{fig:lm_vel_2}
\end{figure*}

\begin{figure*}
    \centering
    \includegraphics[angle=90,width=0.92\textwidth,page=1]{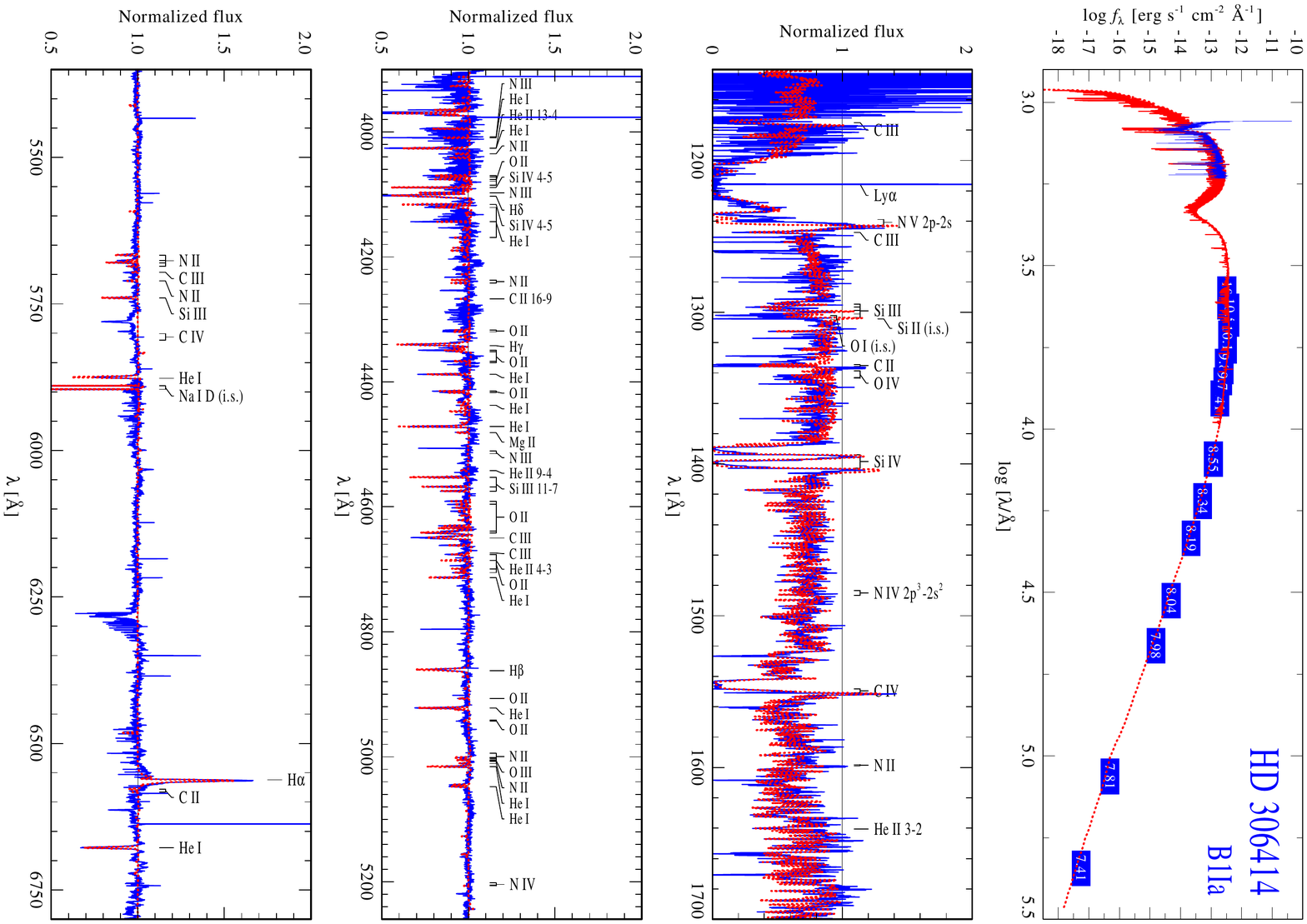}
    \caption{Same as Fig.\,\ref{fig:hd153919}, but for HD\,306414.
    }
    \label{fig:hd306414}
\end{figure*}

\begin{figure*}
    \centering
    \includegraphics[angle=90,width=0.92\textwidth,page=1]{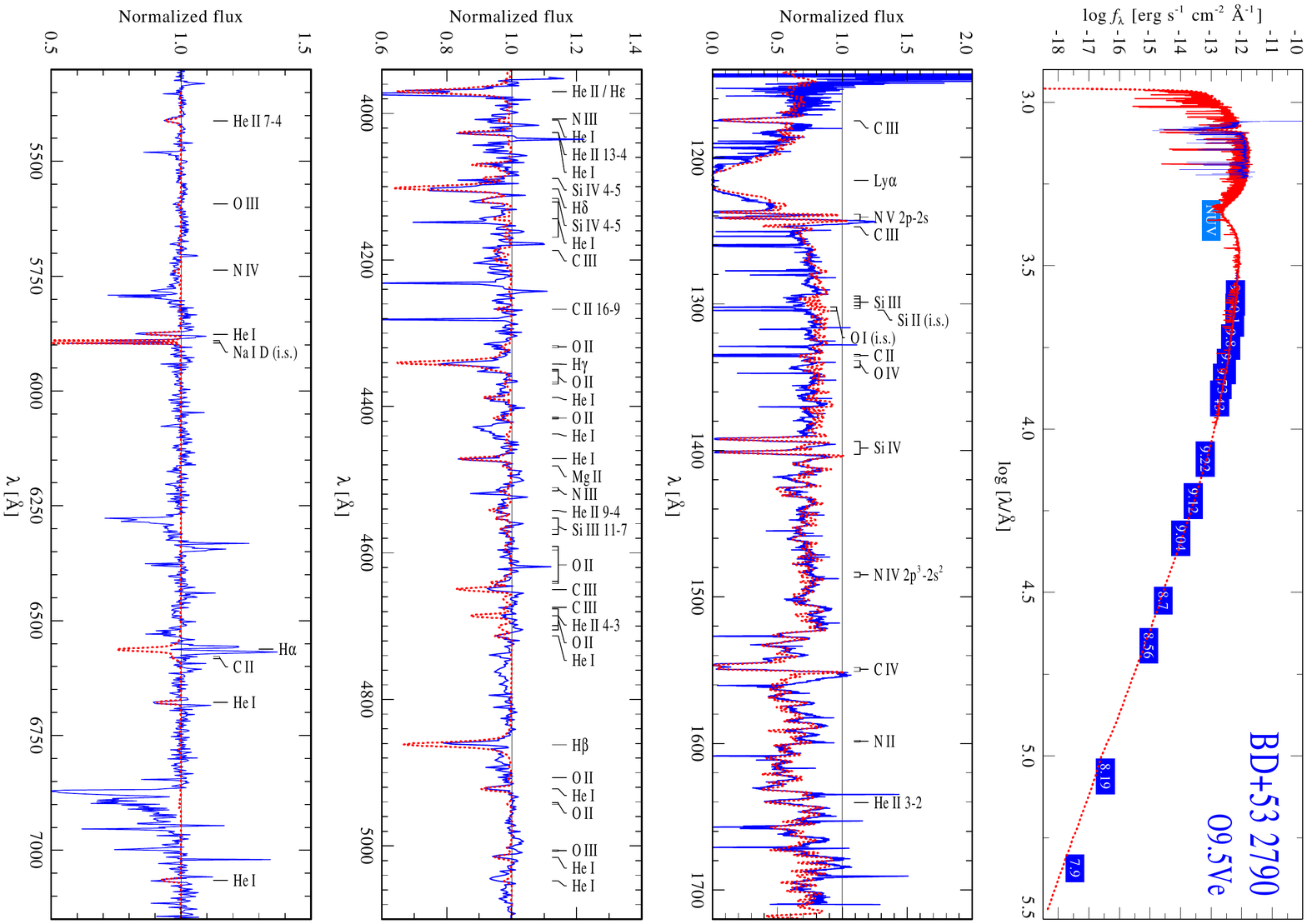}
    \caption{Same as Fig.\,\ref{fig:hd153919}, but for BD+53\,2790. The IR spectrum is clearly dominated by emission from the decretion disk.
    }
    \label{fig:bd+532790}
\end{figure*}

\clearpage

\addtocounter{figure}{-1}

\begin{figure*}
    \centering  
    \includegraphics[angle=90,width=0.92\textwidth,page=2,trim=400 0 0 0,clip]{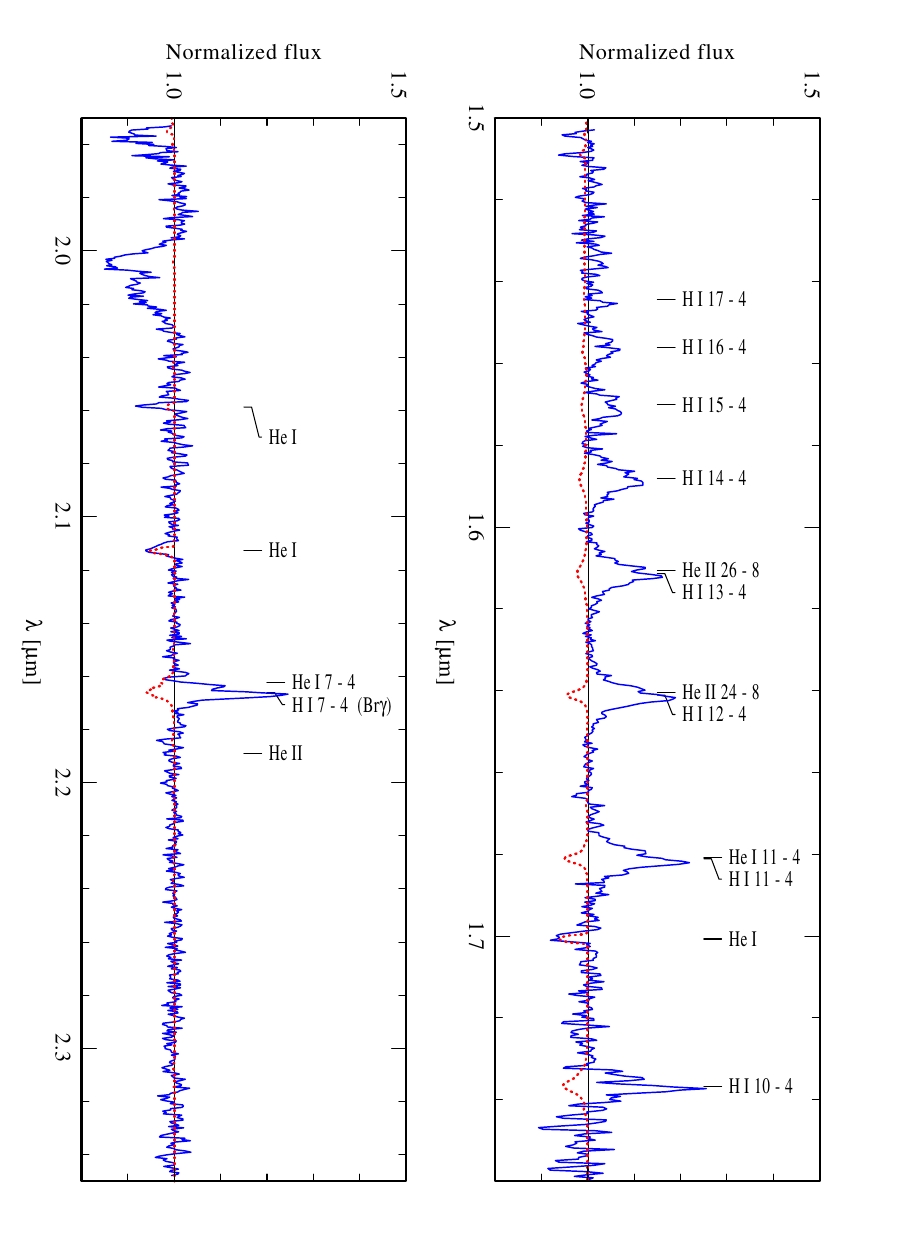}
    \caption{continued. }
\end{figure*}

\begin{figure*}
    \centering
    \includegraphics[angle=90,width=0.92\textwidth,page=1]{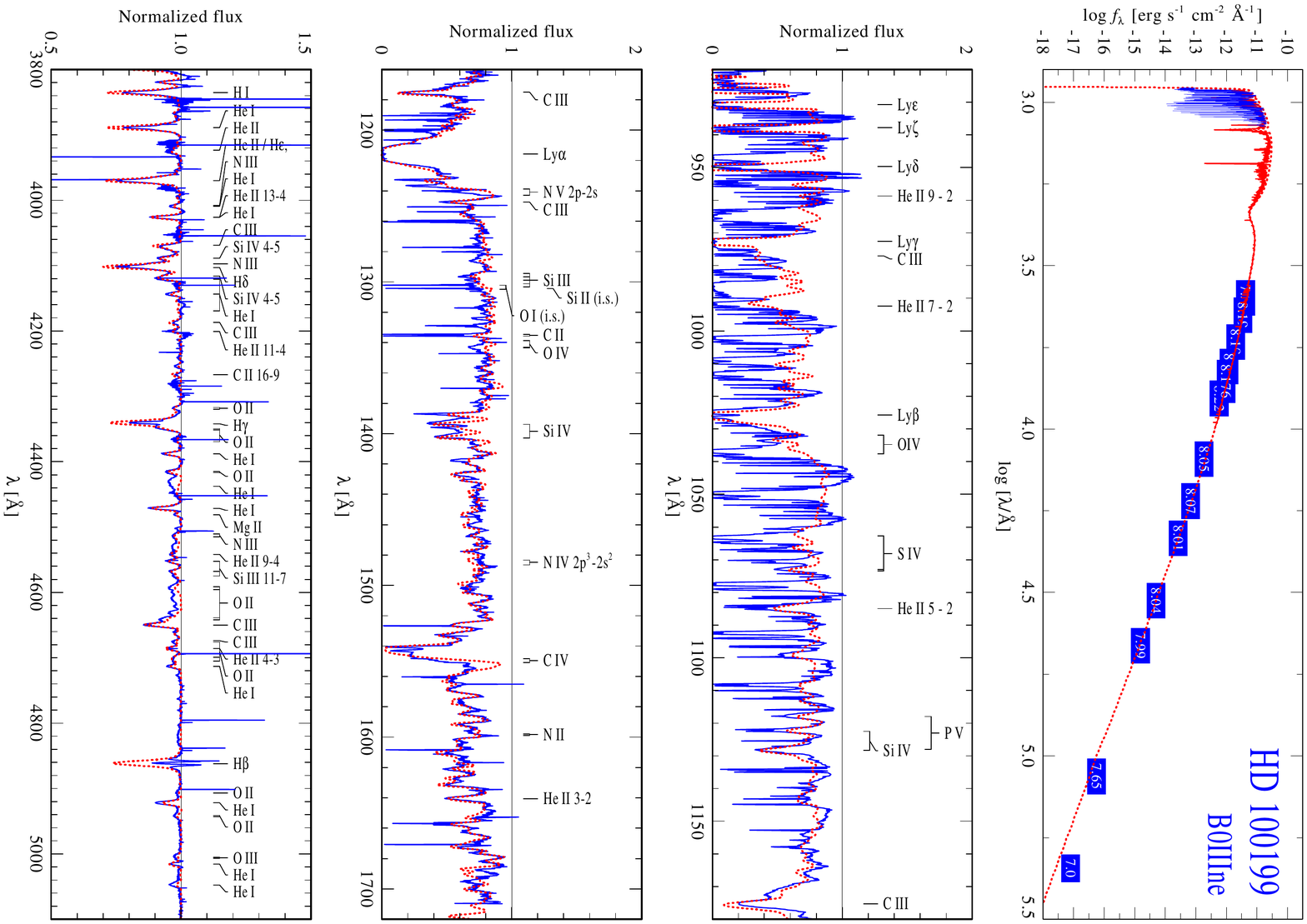}
    \caption{Same as Fig.\,\ref{fig:hd153919}, but for HD\,100199.
    }
    \label{fig:hd100199}
\end{figure*}

\clearpage

\addtocounter{figure}{-1}

\begin{figure*}
    \centering  
    \includegraphics[angle=90,width=0.92\textwidth,page=2,,trim=590 0 0 0,clip]{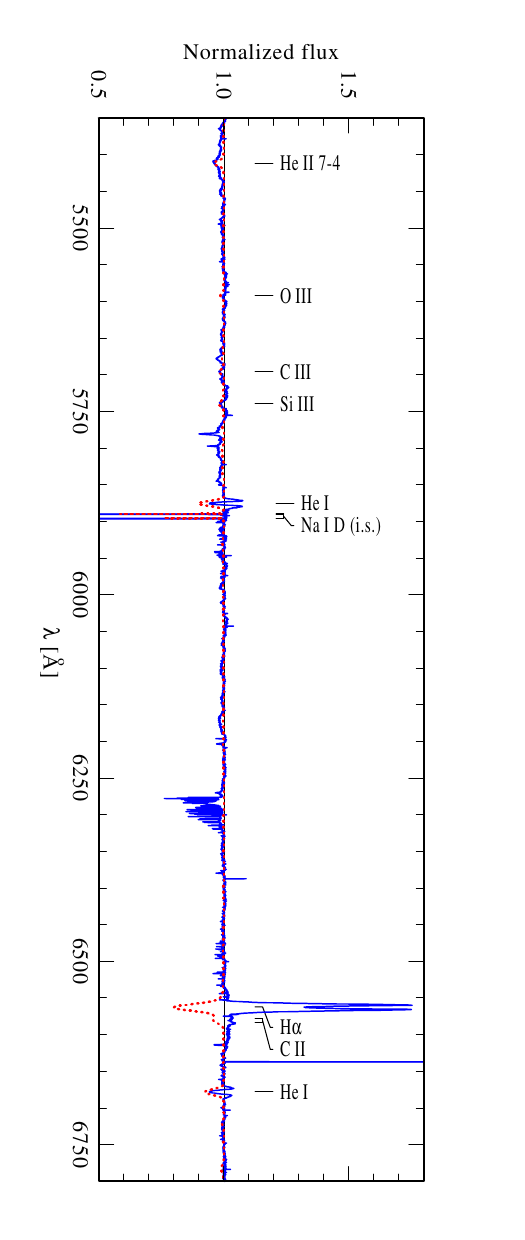}
    \caption{continued.}
\end{figure*}

\end{appendix} 

\end{document}